\documentclass[10pt,preprint]{aastex}
\usepackage{amsmath,amssymb}

\slugcomment{ApJ, 567, 828 (2002); e-print astro-ph/0103333}

\def\ssecref#1{\hbox{\S\,\ref{#1}}}
\def\Secref#1{Section~\ref{#1}}
\def\secref#1{Sec.~\ref{#1}}
\def\apref#1{Appendix~\ref{#1}}

\def\exref#1{(\ref{#1})}

\def\eqref#1{Eq.~(\ref{#1})}

\def\figref#1{Fig.~\ref{#1}}

\def\tabref#1{Table~\ref{#1}}

\def\where{\quad{\rm where}\quad}
\def\const{{\rm const}}

\def\bea{\begin{eqnarray}}
\def\eea{\end{eqnarray}}

\def\phi{\varphi}

\def\({\left(}
\def\){\right)}
\def\[{\left[}
\def\]{\right]}
\def\<{\left\langle}
\def\>{\right\rangle}
\def\l{\left}
\def\r{\right}

\def\d{\partial}
\def\dt{{\d_t}}
\def\diff{{\rm d}}

\def\igamma{\beta}

\def\kappaLL{\kappa_{LL}}
\def\kappaNN{\kappa_{NN}}

\def\vn{{\bf n}}
\def\tkappa{\tilde\kappa}
\def\Idk{{S_d\over(2\pi)^d}\int_0^\infty{\rm d}k}
\def\intk{\int{{\rm d}^d k\over(2\pi)^d}\,}
\def\intkp{\int{{\rm d}^d k'\over(2\pi)^d}\,}
\def\intkpp{\int{{\rm d}^d k''\over(2\pi)^d}\,}
\def\msk{\overline{k^2}}
\def\lambdamax{\lambda_{\rm max}}

\def\intx{\int{\rm d}^d x}

\def\vx{{\bf x}}
\def\vy{{\bf y}}
\def\vk{{\bf k}}
\def\vu{{\bf u}}
\def\vB{{\bf B}}

\def\teddy{\tau_{\rm eddy}}

\def\ueddy{u_{\rm eddy}}
\def\vth{v_{\rm th}}
\def\lmfp{\ell_{\rm mfp}}

\def\kd{k_{\nu}}
\def\kres{k_{\eta}}
\def\yres{y_{\eta}}
\def\gres{\gamma_{\eta}}
\def\Re{{\rm Re}}
\def\Rm{{\rm R}_m}
\def\Pr{{\rm Pr}}

\def\vlp{\mathopen{\boldsymbol{(}}}    
\def\vrp{\mathclose{\boldsymbol{)}}}   
\def\vblp{\mathopen{\boldsymbol{\bigl(}}} 
\def\vbrp{\mathopen{\boldsymbol{\bigr)}}} 

\begin{document}

\title{SPECTRA AND GROWTH RATES OF FLUCTUATING MAGNETIC FIELDS 
IN THE KINEMATIC DYNAMO THEORY WITH LARGE MAGNETIC PRANDTL NUMBERS}
\author{Alexander~A.~Schekochihin\altaffilmark{1}}
\altaffiltext{1}{Present address: Imperial College, 
Blackett Laboratory, Prince Consort Rd, London~SW7~2BW,~U.K.}
\affil{Plasma Physics Laboratory, Princeton University, 
P.~O.~Box~451, Princeton, New~Jersey~08543} 
\email{sure@pppl.gov} 
\author{Stanislav~A.~Boldyrev}
\affil{Institute for Theoretical Physics, University of California, 
Santa Barbara, California 93106}
\email{boldyrev@itp.ucsb.edu}

\and

\author{Russell~M.~Kulsrud}
\affil{Princeton University Observatory, 
Peyton Hall, Princeton, New Jersey 08544}
\email{rkulsrud@astro.princeton.edu}
\date{\centerline{6 March 2001; revised 21 September 2001}}

\begin{abstract}

The existence of a weak galactic magnetic field has been repeatedly 
confirmed by observational data. The origin of this field has not as yet 
been explained in a fully satisfactory way and 
represents one of the main challenges of the astrophysical dynamo 
theory. In both the galactic dynamo theory and the primordial-origin 
theory, a major influence is exerted by the small-scale 
magnetic fluctuations. 
This article is devoted to constructing 
a systematic second-order statistical theory of such small-scale fields. 
The statistics of these fields are studied 
in the kinematic approximation and for the case of large 
Prandtl numbers, which is relevant for the galactic and protogalactic 
plasma. The advecting velocity 
field is assumed to be Gaussian and short-time correlated. 
Theoretical understanding of this kinematic dynamo model
is a necessary prerequisite for any prospective nonlinear dynamo theory. 
The theory is developed for an arbitrary degree of compressibility 
and formally in $d$~dimensions, which generalizes the previously 
known results, elicits the structure of the solutions, and 
uncovers a number of new effects. 
The magnetic energy spectra are studied 
as they grow and spread over scales during the initial 
stage of the field amplification. Exact Green's-function 
solutions are obtained. 
The spectral theory is supplemented by the study of 
magnetic-field correlation functions in the configuration 
space, where the dynamo problem can be mapped onto a particular 
one-dimensional quantum-mechanical problem.  
The latter approach is most suitable for the description 
of the kinematic dynamo in the long-time limit, i.e. when 
the magnetic excitation has spread over all scales present 
in the system.  
A simple way of calculating the 
growth rates of the magnetic fields in this long-time limit 
is proposed. 

\end{abstract}


\section{INTRODUCTION}

\subsection{Astrophysical Motivation}
\label{astro_motives}

The question of the origin of the galactic magnetic field 
has long been a subject of much interest in plasma 
astrophysics. The existence of galactic magnetic fields 
was first inferred by \citet{Alfven-a,Alfven-b} and \citet{Fermi} 
from the properties of cosmic rays. This was later confirmed 
by observational data \citep{Hiltner,Hall_Mikesell}. 
Modern observations indicate that our Galaxy possesses 
a magnetic field that has a large-scale component that is several~$\mu$G 
strong and is coherent on the scales of up to 
a kiloparsec ($\sim 10^{21}$~cm). 
These scales are intermediate between the diameter of 
the galactic disk ($\sim10$~kpc) and 
its thickness ($\sim100$~pc). 
Observations indicate that magnetic fields of similar magnitudes 
(which correspond to magnetic-energy densities comparable to those 
of the fluid motions of the interstellar matter) 
and spatial coherence are common in other galaxies as well 
\citep[see reviews by][]{Kronberg,Beck_etal,Zweibel_Heiles}. 
Most of the theories that have been advanced to explain 
the presence of these fields have in one way or another connected 
their origin with the dynamo action of the interstellar 
turbulence.

The interstellar medium~(ISM) consists of a partially ionized plasma that 
is regularly stirred on the scale of about $100$~parsec by the 
shock waves generated by supernova explosions. The Reynolds number 
of the~ISM is of the order of~$10^5$, which allows 
a fully developed Kolmogorov-type 
turbulent cascade to be set up. The energy generated at 
the outer scale of $\sim100$~parsec is thus transfered approximately 
four decades down to the Kolmogorov inner scale, where 
it is dissipated by the molecular viscosity. 
If an initial seed magnetic field is introduced into such 
a medium, the turbulent velocity field should be expected 
to stretch the magnetic-field lines and thus amplify the field 
via the usual fast-dynamo mechanism \citep{Sakharov,Vainshtein_Zeldovich}. 
The turbulent nature of the physical processes involved 
clearly necessitates a {\em statistical description.}

Traditionally, the hopes for a theoretical explanation of 
the galactic magnetic field have focused upon the 
{\em mean-field dynamo theories} \citep{Parker_alpha,Braginskii_alpha,Steenbeck_Krause_Raedler,Moffatt,Parker_book,Ruzmaikin_Shukurov_Sokoloff}. 
The essential idea is 
to follow the evolution of the volume-averaged (i.e., effectively, 
large-scale) magnetic field subject to two main assumptions. 
First, it is assumed that the initial seed field is small 
and thus, during the initial stage of its evolution, the magnetic 
field is too weak to exert a significant amount of back 
reaction on the hydrodynamic motions. The Lorentz 
forces can therefore be neglected and magnetic field considered 
{\em passive}, which gives rise to the so-called 
{\em kinematic approximation.} 
Second, the large-scale mean field is assumed to be much stronger 
than the small-scale magnetic fluctuations.\footnote{This assumption 
can be circumvented for the Gaussian $\delta$-correlated velocity 
field discussed in \ssecref{kin_dyn_model} 
\citep{Vainshtein_alpha,KA,Boldyrev_alpha}.} 
It turns out that, if these two assumptions hold, and 
if the interstellar turbulence lacks mirror invariance, 
the mean field will grow exponentially 
at a rate proportional to the amount of helicity possessed by 
the turbulent medium. Such mean-field amplification has come 
to be referred to as {\em the $\alpha$~effect.} 

It must, however, be appreciated that the mean-field theory 
faces serious challenges to its theoretical and physical foundations. 
It was pointed out already by \citet{Batchelor_vort_analog} 
that the initial amplification of the magnetic field in the kinematic 
regime would be accompanied by the transfer of the magnetic energy 
to small (nonhydrodynamic) scales. A wide range of such scales 
exists as a result of the huge disparity between the hydrodynamic and 
the magnetic Reynolds numbers of the ISM: while~$\Re\sim10^5$, 
its magnetic counterpart can be as high as~$\Rm\sim10^{19}$. 
The ratio of these two numbers, commonly referred to as the 
magnetic Prandtl number, $\Pr\sim\Rm/\Re$, determines the width of the 
scale interval between the Kolmogorov inner scale~$\kd$, where 
the viscous dissipation cuts off the hydrodynamic turbulence spectrum, 
and the resistive scale~$\kres$, where magnetic energy is dissipated 
by the Ohmic resistive-diffusive mechanism. 
Since $\kres/\kd\sim\Pr^{1/2}$, the subviscous scales accessible 
to the small-scale fluctuating magnetic fields extend over 
as many as seven decades in the wave-number 
space.\footnote{In this context, it should be noted that 
resolving such a broad range of scales in a numerical simulation 
is all but impossible, so theoretical understanding of 
the physics of the small-scale fields is indispensable.} 
In the kinematic 
approximation, the rates of the exponential growth of the small-scale 
magnetic-fluctuation energy and of its transfer toward the small 
scales turn out to greatly (by the factor of~$\sim10^4$) 
exceed the growth rate of the mean field due to 
the $\alpha$-effect, which operates on a time scale associated with 
the overall galactic rotation. 
Unlike the mean field, the magnetic energy 
grows regardless and independently of whether 
the turbulent velocity field has a helical component. 
Thus, the energy 
of the small-scale magnetic fluctuations must be expected to grow 
to equipartition with that of the smallest eddies of the turbulent 
velocity field much faster than the mean field can reach any appreciable 
values. The validity of the kinematic mean-field theory will 
therefore break down long before the growth mechanism predicted by it 
has time to manifest itself \citep{KA}.

The mean-field approach is an attempt to explain the large-scale 
galactic magnetic field in terms of a coherent volume-averaged 
field being amplified by the hydrodynamic helicity effect. 
Besides the impossibility to neglect 
the small-scale fields, there exists another physical consideration 
that makes justifying this view rather problematic. 
Since no magnetic-energy dissipation mechanism 
is available at the large scales, the growth of the large-scale 
field must be consistent with the flux-conservation constraint. 
The most popular theory has been that the magnetic-field lines  
are partly expelled from the galactic 
disk \citep[see][]{Parker_book,Kulsrud_review,Kulsrud_lecture}. 
However, the possibility of flux expulsion has been increasingly 
in doubt \citep{Rafikov_Kulsrud}.  

In view of the fundamental difficulties associated with 
the mean-field approach, one is hard-pressed to seek alternative 
ways to construct an adequate galactic dynamo theory. 
A promising avenue of investigation naturally presents itself 
in the context of the small-scale-field amplification. 
It seems quite reasonably clear and has been an accepted point 
of view since the work of \citet{Batchelor_vort_analog} 
that the growth of the fluctuating fields at subviscous 
scales should culminate in the magnetic-fluctuation energy 
equalizing with that of the smallest turbulent eddies. 
What happens next is a subject of much interest and 
disagreement. The basic point of contention is whether 
the magnetic energy would saturate at viscous \citep{Batchelor_vort_analog} 
or resistive \citep{Vainshtein_Cattaneo,Gruzinov_Diamond-a,Gruzinov_Diamond-b,Gruzinov_Diamond-c} scales, 
or rather proceed to reach full equipartition with 
the turbulence \citep{Biermann_Schlueter}. 

An explanation of the existence of the large-scale galactic magnetic field 
could be within grasp if it were to be demonstrated that, 
once the magnetic-fluctuation energy had equalized 
with that of the smallest eddies, an {\em inverse cascade} would take place, 
forcing the magnetic excitation to keep growing and to gradually move 
towards ever-larger scales until equipartition were achieved at all 
scales up to the energy-containing ones.
A statistical theory of this 
sort would predict the emergence of {\em large-scale magnetic fluctuations}   
with energy comparable to that of the turbulence 
(see \secref{sec_future} for further discussion of these 
matters). From the point 
of view of the mean-field approach, these fluctuations would still have 
a zero statistical average. However, observationally, they would 
locally appear as a field coherent at large 
scales \citep[cf.][]{Blackman_no_alpha}. 

This idea faces a serious setback when measured against 
the actual parameters of the~ISM. Namely, the typical 
energy-containing scales of the galactic turbulence, to which 
one can expect the inverse cascade to bring the magnetic energy, 
are of the order of 100~parsec (the average distance between the 
supernovae, which drive the turbulence), or about 10~times smaller 
than the coherence scale of the observed large-scale galactic magnetic 
field. This problem can be resolved if one 
follows \citet{Kulsrud_etal_proto} 
in their recent suggestion to consider the possibility of 
a {\em primordial} origin of the galactic magnetic field. 
The primordial-origin hypothesis regards the large-scale galactic 
field not as a product of {\em galactic} turbulent dynamo mechanisms,  
but as a residue of the analogous processes that take place in 
{\em protogalaxies} before they collapse into 
galaxies.

Indeed, the conditions in the protogalactic plasma cloud seem 
to be more favorable than those in the galactic~ISM for the operation of 
a bona~fide turbulent dynamo. As does the~ISM, the protogalactic 
plasma supports a Kolmogorov-type turbulence, this time driven 
by the shock waves originating from the instabilities associated 
with the forces of gravitational collapse. These instabilities 
occur on the scales comparable to the size of the protogalaxy
($\sim100$~kiloparsec) and thus, unlike galaxies, protogalaxies 
know no disparity between the system size and the energy-containing 
scale of the turbulence. The hydrodynamic Reynolds number 
of the protogalactic plasma is~$\Re\sim10^4$, so an inertial range 
about three decades wide is available for the Kolmogorov turbulent 
energy cascade. Numerical simulations indicate  
that such a cascade is indeed set up \citep{Kulsrud_etal_proto}. 
The protogalactic Prandtl 
number is even larger than the galactic one: $\Pr\sim10^{22}$, 
which allows as many as eleven decades of subviscous scales 
accessible to the magnetic fields. 

The seed magnetic fields in the protogalaxy could be  
created from an initial state with no fields at all 
by the so-called Biermann-battery 
mechanism \citep{Biermann_battery,Kulsrud_etal_proto}. 
This battery action is associated with the thermoelectric term in 
the plasma Ohm's law. This term is non-zero provided 
the motions are nonbarotropic ($\nabla p\times\nabla\rho \ne 0$).
Since the violent beginnings of the protogalaxy may be assumed 
to allow nonbarotropic large-scale pressure fluctuations, 
it can be estimated that the battery is capable of generating 
large-scale magnetic fields of about~$10^{-21}$~G before the dynamo 
term becomes dominant. 

The magnetic fields are then amplified 
and carried over to small scales by the same \citet{Batchelor_vort_analog} 
mechanism that foiled the galactic 
mean-field theory. After the equipartition with the smallest turbulent 
eddies is reached, an inverse cascade may be envisioned that 
brings the magnetic energy back to the large scales. 
As the protogalaxy collapses into the galactic disk, its magnetic 
field is compressed and becomes the initial magnetic field 
of the newly formed galaxy. If the pre-collapse protogalactic 
magnetic field is amplified by the protogalactic turbulence 
to sufficiently large values, it may well be enough to give rise, 
upon compression, to a galactic field with the strength and scale 
of coherence reasonably close to those observed 
\citep{Howard_Kulsrud,Kulsrud_etal_report,Kulsrud_review,Kulsrud_lecture}. 
[A midway approach is to consider the residual 
field resulting from the {\em protogalactic} dynamo as the seed field 
for further operation of the {\em galactic} dynamo mechanisms such 
as, e.g, the $\alpha$-$\Omega$~dynamo 
\citep{Parker_book,Kulsrud_review,Kulsrud_lecture} 
\citep[see also][]{Blackman_no_alpha}.] 
While this scenario is certainly largely hypothetical, 
it offers a physical possibility that can hardly be neglected. 

In all of the above discussion, the fluctuating magnetic fields     
in general and the small-scale fields in particular have played 
a prominent role. 
Under additional assumptions about 
the turbulent velocity field, the kinematic stage of their evolution 
can be treated analytically and exactly. {\em Any theoretical 
understanding of the processes that occur once the nonlinear 
stage is initiated, must rely on 
a solid statistical theory of the small-scale 
kinematic dynamo.} 
In this paper, we give a detailed and systematic exposition 
of the small-scale kinematic dynamo theory, 
following an often-neglected Cartesian principle: 
``\dots je me persuadai\,\dots que, pour toutes les opinions 
que j'avais re\c{c}ues jusques alors en ma cr\'eance, je ne pouvais 
mieux faire que d'entreprendre une bonne foi de les en \^oter, afin 
d'y en remettre par apr\`es, ou d'autres meilleures, ou bien 
les m\^emes, lorsque je les aurais ajust\'ees au niveau de 
la raison''\citep{Descartes}.\footnote{``\dots I became 
convinced\,\dots 
that the best I could do with all the notions that I had thitherto 
taken into my credence was to have the good sense of ridding my mind 
thereof, in order to subsequently put back~in either other, better ones, 
or even the same, once I had arranged them according to reason'' 
(Descartes, {\em Discourse on Method}).}

\placetable{tab_params}
\notetoeditor{This table can be placed anywhere in the above subsection}

\subsection{The Kinematic Dynamo Model}
\label{kin_dyn_model}

The full-scale nonlinear theory of the astrophysical dynamos 
is, of course, a formidable challenge and, in practical terms, 
a degree more complicated than even the problem of fluid turbulence. 
One must therefore try to find simplified 
models in order to be able to extract any theoretical understanding 
of the magnetic dynamo problem. The kinematic stage of the 
magnetic-field growth, when the field is so weak that it exerts 
no appreciable back force on the medium, has traditionally been 
the focus of theoreticians' attention. Indeed, if the velocity field 
is decoupled from the magnetic field and can be fed into the 
problem in some given form, the induction equation becomes formally 
linear (in the magnetic field) and the task of solving it starts to 
appear within theory's reach. 
Thus, we consider the magnetic field evolving according to the 
induction equation (formally in $d$~dimensions):
\bea
\label{induction_eq}
\dt B^i = - u^k B^i_{,k} + u^i_{,k} B^k - u^k_{,k} B^i 
+ \eta\Delta B^i,
\eea
where $u^i(t,\vx)$ is the externally specified 
advecting velocity field, 
$u^i_{,k}=\d u^i/\d x^k$, $B^i_{,k}=\d B^i/\d x^k$, and 
the Einstein summation over repeated indices is used throughout. 
Two basic avenues of 
research have been pursued in the literature on the kinematic 
dynamo problem. One of them is to study the magnetic-field 
amplification in some prescribed {\em deterministic}, though possibly chaotic 
(i.e.~{\em stationary and stochastic in space}), 
fluid flows and/or in specific geometries 
\citep[see, e.g.,][]{Moffatt,Ruzmaikin_Shukurov_Sokoloff,Stretch_Twist_Fold}.
Another approach is 
to consider {\em random (in time)} velocity fields with prescribed 
statistics and try to determine the resulting statistics 
of the magnetic field \citep{Kazantsev,KA}. 
As the dynamo models of the latter kind 
appear to better reflect the turbulent nature of the fluid motions 
that occur in the astrophysical applications of interest to 
us (ISM and protogalaxies: see~\ssecref{astro_motives}),
we restrict ourselves to such statistical line of research only. 
Additional assumptions of spatial homogeneity and isotropy of the 
system and statistical stationarity of the velocity field 
are usually made. 
Furthermore, we assume mirror-invariance, thus excluding 
the helicity effects, which are not expected to be felt 
at the time scales relevant to the small-scale-field physics 
\citep[see discussion in~\ssecref{astro_motives} above and in][]{KA}.

Unfortunately, the kinematic assumption alone is not sufficient to 
turn the turbulent dynamo problem into a tractable one. The realistic 
turbulent velocity fields possess nontrivial 
intermittent statistics, which are hard to handle. 
It is necessary to assume something about them, fully 
appreciating the fact that any such assumption is bound to 
be highly artificial. An obvious such assumption, to which 
many turbulence theories gravitate, is to consider a {\em Gaussian}
velocity field. This brings about an enormous simplification  
due to the splitting property of the Gaussian averages.  
However, even the Gaussian assumption does not fully 
remove the usual turbulence moment closure problem. 
The remaining complication has to do with the fact that, in general, 
the advecting flow is finite-time correlated. 

When the advecting velocity field has a finite correlation 
time, the equations for statistical quantities such as 
the moments of the magnetic field 
cannot be obtained in a closed form. 
Moreover, introducing a finite velocity correlation 
time into the problem leads to sensitive dependence 
of the resulting statistics on the specific structure 
of the velocity correlations in time and space. 
[For more detailed discussion of these issues and 
further results with regard to the finite-correlation-time 
effects, we address the reader to papers 
by \citet{vanKampen_review,Boldyrev_tcorr,SK_tcorr}, 
and references therein.]
In order to obtain a solvable kinematic dynamo model,  
we must therefore take another drastic 
simplifying step. This consists in 
assuming that the velocity field, besides being Gaussian, also 
possesses {\em the white-noise property,} i.e., it is a random process 
{\em $\delta$-correlated in time:} 
\bea
\label{velocity_corr}
\<u^i(t,\vx)u^j(t',\vx')\> = \delta(t-t')\kappa^{ij}(\vx-\vx').
\eea 
Such a 
synthetic velocity field is sometimes called {\em the Kraichnan ensemble,} 
in deference to the author who first proposed it as a model 
for studying the passive-scalar problems such as the advection 
of temperature, or of concentration of 
an admixture \citep{Kraichnan-1,Kraichnan-3}. In application 
to the magnetic fields, this model was proposed independently 
by \citet{Kazantsev}, whose work, in fact, predates 
Kraichnan's. In physical terms, the Kazantsev--Kraichnan assumption amounts 
to the short-correlation-time approximation and is valid if 
the advecting velocity field is correlated at times much 
smaller than the characteristic time of the dynamo action. 
The latter is of the order of the inverse velocity gradient 
(the {\em eddy-turnover time} of the velocity field). 

When the white-noise assumption is put in force, the closure 
problem resolves itself. Thus, one of the main reasons for 
the Kazantsev--Kraichnan passive-advection paradigm 
being so attractive 
is that it is one of the few available ``Ersatz-models'' of turbulence, 
which hold promise of exact solvability.
However, the existing theory is not complete, 
and the statistics of passive advection continue to generate 
considerable interest, for the problem of passive advection can 
serve as a vehicle for developing a set of analytical tools 
appropriate for attacking more realistic turbulence models.
 
The objective of this paper will be to develop a theory 
that would allow us to calculate, under the assumptions 
explained above, {\em the two-point second-order correlation 
function} of the advected field~$B^i$:
\bea
\label{mfield_corr}
\<B^i(t,\vx)B^j(t,\vx')\> = H^{ij}(t,\vx-\vx').
\eea
Here and everywhere in this work, the angle brackets denote 
ensemble averaging with respect to the random 
advecting velocity field $u^i$, as well as with respect to the 
initial distribution of $B^i$. The latter, however, should not 
greatly affect matters after a finite transient time has elapsed 
(physically, about one eddy-turnover time). 
The two-point, second-order correlation functions have a clear 
physical interpretation: their Fourier transforms are 
{\em the energy spectra.}

\subsection{General Scenario of the Small-Scale-Field Dynamo}
\label{outline}

The study of the kinematic dynamo problem in the above formulation 
was pioneered by \citet{Kazantsev}. 
Kazantsev reduced the problem of finding the two-point correlation 
functions of the fluctuating magnetic fields to solving a certain 
quantum mechanical problem that described a particle with variable 
(position-dependent) 
mass moving in a one-dimensional potential well. The eigenfunctions 
and the energy levels of the particle's Hamiltonian corresponded 
to the magnetic-field correlation functions with stationary spatial 
profiles and their overall growth rates, respectively. 
The ground state of the particle 
gave rise to the largest growth rate and thus determined the long-time 
evolution of the magnetic fluctuations.  
Importantly, the exact shape of the potential 
depended on the particular form of the 
velocity correlation functions.\footnote{Recently it was demonstrated 
that statistics of general passive tensor fields, such as gradients 
of temperature or concentration, tensor products of vectors, etc., 
can also be reduced to a quantum mechanical problem. Remarkably, 
this problem is exactly solvable and describes $d$ particles on 
the line which interact pairwise with the potential~$\sim 1/(x_i-x_j)^2$, 
the so called Calogero-Sutherland 
potential \citep{Bernard_Gawedzki_Kupiainen,BS_metric}.} 

In order to extract the general 
properties of the problem, Kazantsev concentrated his attention on 
the fluctuating magnetic fields at scales very small compared 
to the correlation length of the velocity field. 
In this limit, 
the two-point velocity correlation functions, which depend 
on the distance between the points, can be expanded around 
the origin. Two principal subclasses of the two-point 
kinematic-dynamo problem can be identified in this context, 
depending on the physical setting one is interested in. 
If a problem with a large magnetic Prandtl number is considered, 
i.e., if the resistive-diffusive scale~$\kres^{-1}$ is much smaller 
than the characteristic scale~$\kd^{-1}$ of the advecting flow, 
viz.,~$\kres/\kd \sim \Pr^{1/2} \gg 1$, 
we find ourselves dealing with 
what is customarily referred to in the literature 
as {\em the Batchelor regime} [after the original 
work of \citet{Batchelor_regime} 
where this view of passive advection was first elaborated]. 
In this case, the Taylor expansion of the 
velocity correlator~$\kappa^{ij}(\vy)$ can be used, 
whose lowest-order $\vy$-dependent term is quadratic 
(see \apref{ap_review}). Since in the astrophysical applications 
that are of most interest to us (the interstellar medium and 
the protogalactic plasma), the Prandtl number is very large, 
it is the Batchelor regime that we shall concentrate on in this work.
The models with small Prandtl numbers, where the scale ratio 
between the flow and the magnetic field is reversed, 
$\kres/\kd \sim \Pr^{1/2} < 1$, lead to nonanalytic $\vy$~dependence 
for the velocity correlator that magnetic fluctuations 
``feel'' at small scales. Slightly different quantum mechanics 
result \citep{Kazantsev,Vainshtein_SSF-b,Vainshtein_SSF-c,Kichatinov,Vainshtein_Kichatinov,Vergassola,Rogachevskii_Kleeorin}. 
These matters fall beyond the scope of this paper and 
will be reported elsewhere.

\begin{figure}
\plotone{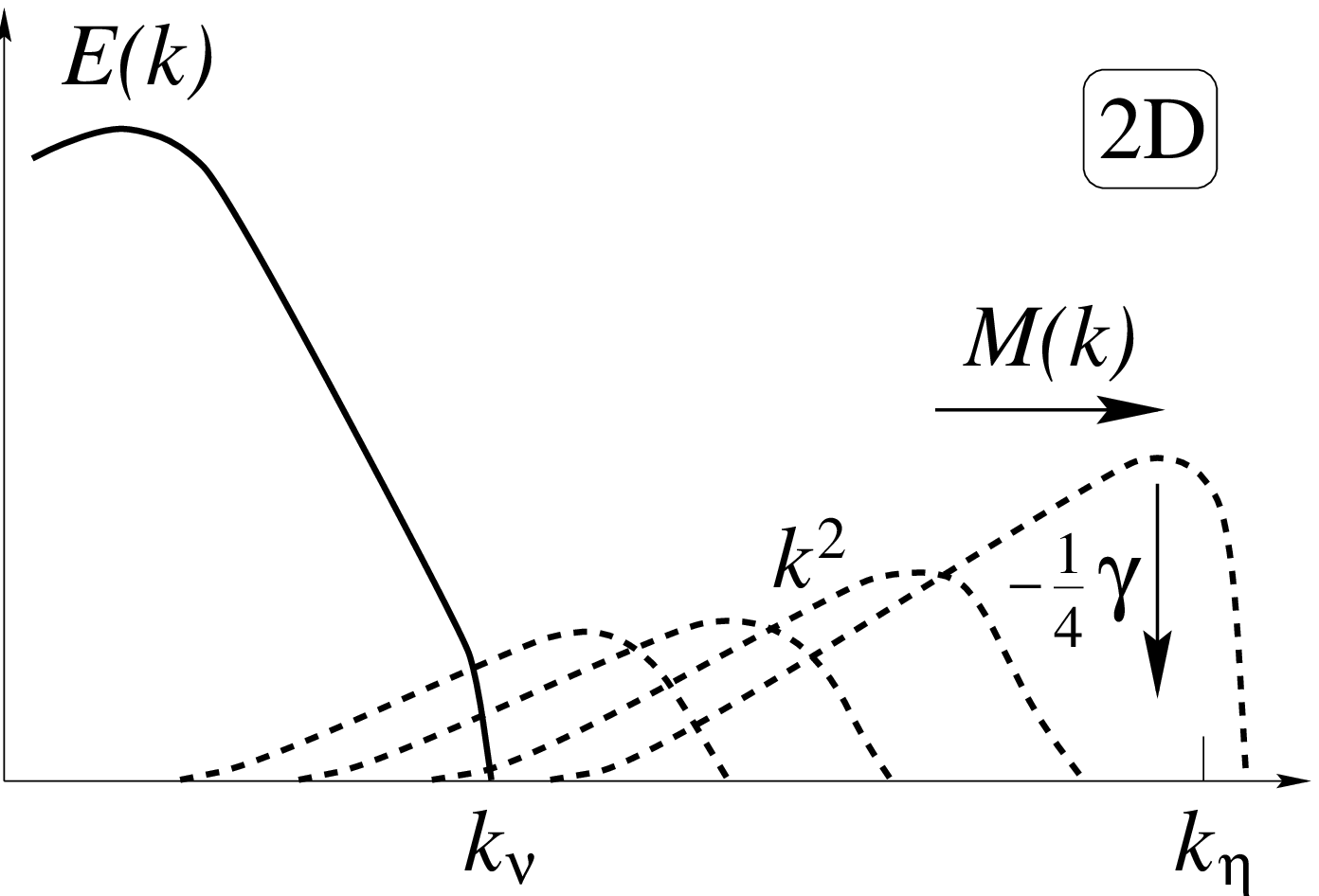}
\caption{Sketch of the small-scale dynamo in~2D. The number of 
excited modes grows, while each individual mode decays. Both processes 
occur exponentially fast. The total growth rate of the magnetic 
energy {\em in the diffusion-free regime} is~$2\gamma$. 
The per-mode decay rate and the spectral index depend 
on the degree of compressibility. 
The numbers on the sketch correspond to the incompressible case.
\label{fig_dynamo2D}}
\end{figure}

\begin{figure}
\plotone{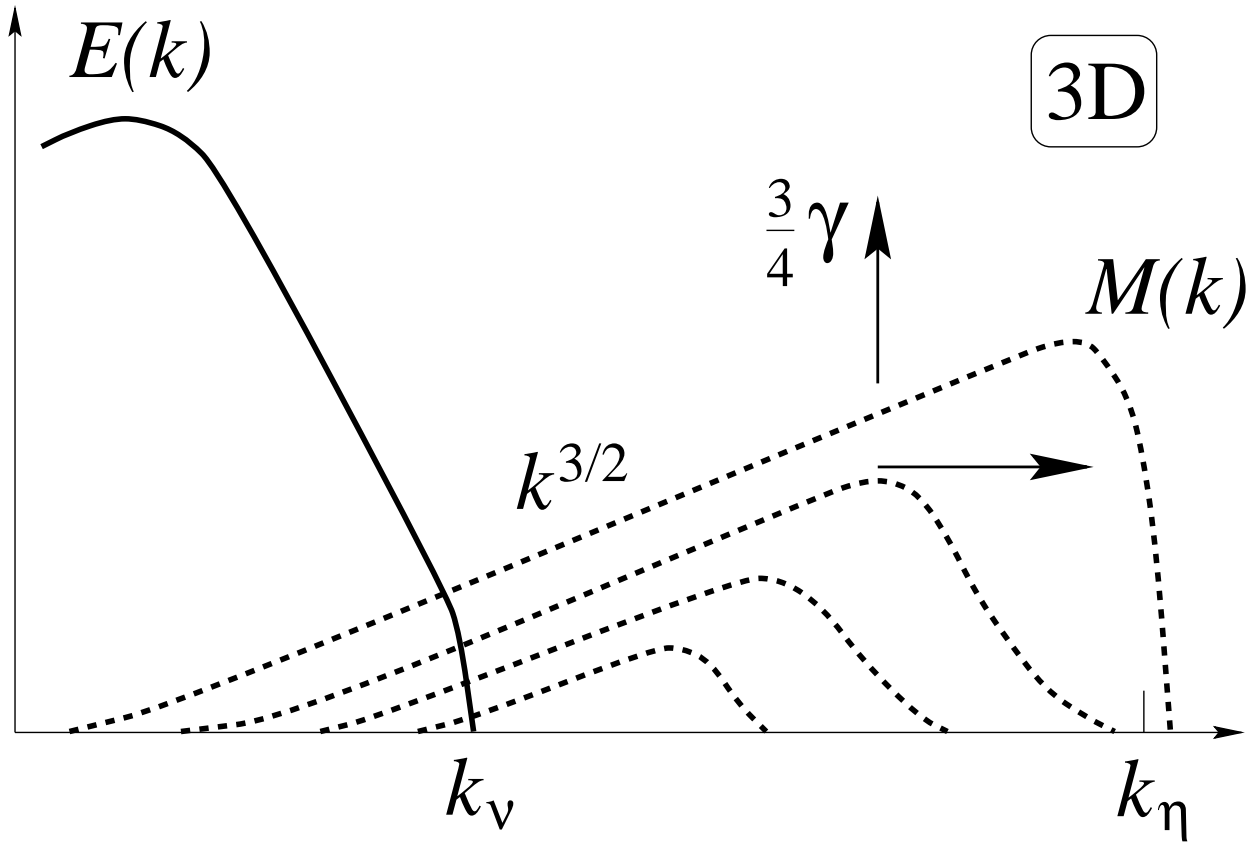}
\caption{Sketch of the small-scale dynamo in~3D. Both the number 
of modes and the amplitudes of all individual modes grow exponentially 
in time. The total growth rate of the magnetic 
energy {\em in the diffusion-free regime} is~$2\gamma$. 
The per-mode growth rate depends on the degree of compressibility, 
but the spectrum is always~$k^{3/2}$. 
The numbers on the sketch correspond to the incompressible case. 
\label{fig_dynamo3D}}
\end{figure}

Let us now outline the general scenario that emerges from 
the investigation of the statistical evolution of small-scale 
fluctuating magnetic fields in the Batchelor 
regime \citep{KA}. 
Let us assume that, initially, the magnetic fields are excited 
at scales comparable to the characteristic 
scale of the advecting flow~$\kd^{-1}$. 
(In the astrophysical context discussed above, this assumption is 
consistent with the current undertanding of the physics 
of seed fields --- see \ssecref{astro_motives} and references therein.)  
Due to the convective and 
line-stretching (``dynamo'') action of the velocity field, 
two processes are initiated and proceed exponentially 
fast in time, at the rates comparable to the eddy-turnover 
rate of the velocity field. 
First, the magnetic fluctuations spread over a widening range 
of scales, their bulk shifting towards ever-smaller ones.
A power-like spectrum forms behind the advancing peak.
Second, the level of excitation (amplitude) of each spatial 
Fourier mode either grows (in~3D) or decreases (in~2D). 
The combination of these two processes results in the exponential 
growth of the total magnetic energy. 
\figref{fig_dynamo2D} and \figref{fig_dynamo3D} illustrate 
the evolution of the magnetic-fluctuation spectra in 
two and three dimensions respectively. 
Note that in ~2D, 
the overall growth of the energy is due entirely 
to the spreading of the excitation over an increasing number 
of modes, while in~3D, each mode also grows by itself. 
We call this stage of the kinematic dynamo {\em the diffusion-free 
regime.} 

\placefigure{fig_dynamo2D}
\placefigure{fig_dynamo3D}
\notetoeditor{These two figires should appear side-by-side in print}

Eventually, the magnetic fluctuations reach 
the diffusive (resistive) scales~$\sim\kres^{-1}$. 
The effect of the Ohmic dissipation is to 
check the decrease of the characteristic scale of the magnetic field. 
The evolution of the magnetic energy is 
now determined exclusively by the exponential growth (in~3D) 
or decay (in~2D) of the level of excitation in each particular mode. 

We will see in our analysis that, while the evolution 
of the magnetic fluctuations starts at scales comparable 
to those of the advecting velocity (henceforth referred to as 
{\em large}, or {\em integral}, scales), practically all of their  
energy will shift to the small scales, $k\gg\kd$, after 
a few eddy-turnover times. Thereupon, the spectral properties 
of the small-scale fields can for the most part be understood 
in disregard of the small fraction of the magnetic energy 
that is situated at the large scales. 
The auxiliary model where the effect of the large scales 
is neglected and all the magnetic energy is 
considered to be concentrated at small scales, 
will also be called {\em the diffusion-free regime} before 
the resistive scales are reached,  
and {\em the resistive regime} afterwards. 
For both cases, exact Green's-function solutions can 
be obtained (see~\ssecref{spreading} and~\ssecref{spreading_diss}). 

The eventual spreading of the excitation over all scales 
available in the system, including the resistive and the large, 
ushers the true long-time asymptotic regime 
(or {\em the long-time limit}), 
which was the principal subject of Kazantsev's investigation. 
In three dimensions, a stationary correlation profile is 
expected to form which, if the Prandtl number exceeds a certain 
critical value, continues to grow exponentially. 
This is the eigenstate arising from Kazantsev's 
quantum mechanics, and Kazantsev's objective was to determine 
the growth rate and the scaling of the correlation function. 

While the analysis of the small-scale asymptotics 
allowed him to constrain this growth rate between certain, 
relatively narrow, bounds, Kazantsev identified the main difficulty 
of the theory: effectively, the potential that appeared in his 
quantum mechanics was of the inverse-square kind ($\sim-1/x^2$) 
and therefore had to be 
regularized at {\em both} small (resistive) {\em and} large (velocity-field) 
scales, in order for the eigenstates of the problem to 
be determined \citep[see, e.g.,][]{Morse_Feshbach}. 
While on the small-scale side the resistive regularization could be 
included exactly,
the boundary conditions at large scales 
were unknown, so there was not enough information to fix the dynamo growth 
rate. A number of authors took up the study of Kazantsev's  
quantum-mechanical model in the 1980's. 
Much mathematical insight was gained into the properties 
of Kazantsev's quantum mechanics, 
but the integral-scales problem was mostly 
circumvented by choosing particular reasonable forms of 
the velocity correlation function \citep{Ruzmaikin_Sokoloff,Novikov_Ruzmaikin_Sokoloff,Artamonova_Sokoloff,Kleeorin_Ruzmaikin_Sokoloff,Maslova_Ruzmaikin}.  

The particular cases analyzed in the literature
indicate that the second-order statistics (energy spectra 
and growth rates) are insensitive to the exact form 
of the large-scale regularization 
\citep[see][ \ssecref{matching}, and \apref{ap_next_order} below]{Novikov_Ruzmaikin_Sokoloff,Artamonova_Sokoloff,Kleeorin_Ruzmaikin_Sokoloff,Maslova_Ruzmaikin,KA,Subramanian-a,Subramanian-b,Subramanian-c}. 
The growth rate in three dimensions can 
be determined exactly and turns out to be slightly smaller 
[by~$\sim1/\(\ln\Pr\)^2$] than the growth rate of each individual mode 
that operated in the diffusion-free regime. 
In two dimensions, the steadily growing bound eigenstates do not exist, 
and a continuous spectrum of eigenvalues 
is formed that is bounded only by zero 
and therefore indicates that the exponential decay of the magnetic 
fluctuations in the resistive regime is replaced by 
a power-like decay in the long-time 
limit \citep{Novikov_Ruzmaikin_Sokoloff}. 
In this context, we recall that, 
as was predicted most generally by Zeldovich's antidynamo 
theorem \citep{Zeldovich_antidynamo}, only transitory growth is 
possible in two dimensions, and the magnetic field must eventually 
be dissipated by the resistivity. 

The goal of this paper is to give a coherent general 
exposition of the two-point second-order statistical theory 
of the kinematic dynamo. The emphasis is naturally placed 
upon the universal aspects of the problem, which emerge 
when the statistics of small-scale fields are considered. 
As long as specific functional features of the velocity 
correlation function~\exref{velocity_corr} remain inessential, 
only its small-scale expansion affects the statistics of  
the passive magnetic fields:
\bea
\kappa^{ij}(\vy) = \kappa_0\delta^{ij} 
-{1\over 2}\,\kappa_2\(y^2\delta^{ij} + 2ay^iy^j\) + \cdots.
\eea
In this case, the kinematic dynamo problem contains only two 
essential dimensionless 
parameters: the dimension of space~$d$ and the degree of 
compressibility~$a$ of the velocity field. Rather than considering 
various particular cases, such as incompressible, $a=-1/(d+1)$, 
irrotational, $a=1$, three-, or two-dimensional dynamo, 
we work in general, explicitly keeping all dependences on these 
parameters. This enables us to construct a solid theoretical 
framework which incorporates all previously available results, 
elicits their underlying structure, and allows us to establish  
several previously unknown features and facets of 
the kinematic dynamo problem. Besides the general evolution equations 
for the magnetic-field spectra and configuration-space correlation 
functions that are derived for arbitrary compressibility and 
spatial dimension [such as equations~\exref{MC_eq}, \exref{SSF_eq}, 
\exref{eq_H_LL}, and~\exref{Kaz_eq}], 
the new results include, for example, the 
exact Green's-function solution for the small-scale-field spectra 
in the resistive regime~[\eqref{M_Macdonald}] and a simple 
regularization procedure that allows to calculate the dynamo 
growth rates in the long-time limit~(\ssecref{matching}).

This paper contains three more Sections and three Appendices. 
In \secref{sec_spectra}, 
the kinematic dynamo problem is studied in its spectral form. 
Time evolution of the magnetic energy and its propagation over 
scales are described up to the point when all available scales are excited:
from those where the velocity field operates to 
those at which the magnetic energy is diffused by the Ohmic 
resistive damping. 
In \secref{sec_corr_fns}, we study the 
two-point correlation functions of the magnetic fluctuations 
in the spirit of Kazantsev's configuration-space quantum-mechanical 
approach, which is convenient for handling the kinematic dynamo 
problem at all scales in the long-time limit. 
Each of these sections starts with a brief summary of its 
internal organization and of the results contained therein. 
\secref{sec_future} contains concluding remarks of physical 
nature. In particular, we discuss the possible directions of 
future research and some strategies whereby 
the yet-to-be-developed nonlinear small-scale-dynamo theory 
could be built upon the foundation laid by the kinematic 
theory worked out in this paper. 
Under \apref{ap_review}, we include a number of technical appendices 
dealing with the basic properties of the correlation functions 
of isotropic fields. 
In~\apref{ap_Novikov}, we state the 
fundamental Gaussian averaging theorem that lies at the core 
of the derivations of closed equations for the correlation 
functions in the Kazantsev--Kraichnan model: the Furutsu--Novikov 
formula \citep{Furutsu,Novikov}. 
Finally, \apref{ap_next_order} contains some overflow asymptotic 
results pertaining to the material of~\secref{sec_corr_fns}.

\section{SPECTRA OF MAGNETIC FLUCTUATIONS}
\label{sec_spectra}

We will start our investigation of the kinematic dynamo by 
looking at the spectra of magnetic fluctuations. While in most of 
the works on the subject \citep{Kazantsev,Ruzmaikin_Sokoloff,MRS_equations,Novikov_Ruzmaikin_Sokoloff,Kichatinov,Vainshtein_Kichatinov,Artamonova_Sokoloff,Kleeorin_Ruzmaikin_Sokoloff,Maslova_Ruzmaikin,Vergassola,Rogachevskii_Kleeorin,Subramanian-a,Subramanian-b,Subramanian-c}, 
the configuration-space view was chosen 
because of a transparent quantum-mechanical form the problem took, 
it is easier to gain a general physical understanding of the processes 
under way if one first gains a firm foothold in the wave-number space. 
The statistical-evolution scenario outlined in the Introduction, 
viz., the growth and spreading of the magnetic excitation over scales, 
will naturally emerge in the $\vk$-space description. 
Another important consideration is that the correlation functions 
are better recognized as such in the Fourier space, while in the 
$\vx$~space, only the functions that are Fourier transforms 
of the proper $\vk$-space correlation functions constitute the class 
of allowed solutions (see \apref{ap_iso_FT}).
 
The evolution equation for the magnetic-field spectrum was  
first derived by \citet{Kazantsev} and, 
in a somewhat different, but equivalent, form, 
by \citet{Kraichnan_Nagarajan}. 
\citet{KA} later developed a rather 
detailed theory of magnetic fluctuation spectra. Here we derive 
some of the existing results in a more general form 
(in $d$~dimensions and for arbitrary degree of compressibility of 
the advecting flow) and then proceed to present several new 
results on the spectral theory of small-scale fields. 
In~\ssecref{mode_coupling}, we derive the general integral 
equation that governs the evolution of the magnetic spectrum. 
From it, we infer the exponential growth of magnetic energy  
and the exponential decrease of the characteristic scale of 
magnetic fluctuations. In~\ssecref{small_scale_fields}, 
we obtain a partial differential equation that 
describes the behavior of the magnetic spectra at scales 
much smaller than those of the advecting flow. In~\ssecref{spreading}, 
we study the exact Green's-function solution for the spectrum 
of the small-scale magnetic fluctuations in the diffusion-free regime. 
In~\ssecref{spreading_diss}, we proceed to the resistive regime. 
The previously unknown exact Green's-function solution is 
obtained in quadratures that is valid simultaneously 
in both inertial and resistive scale ranges. 
In~\ssecref{quasist_k}, we take up 
the question of the long-time asymptotics of the magnetic spectra 
that emerge after both resistive and integral scales 
have been excited. The latter problem is better suited to be treated 
in the configuration space. This brings us to the material 
presented in~\secref{sec_corr_fns}.

\subsection{The General Equation for the Magnetic-Energy Spectrum}
\label{mode_coupling}

We shall consider the kinematic dynamo problem 
as a problem of Gaussian passive advection 
defined by equations~\exref{induction_eq} and~\exref{velocity_corr}.
The velocity correlation function in the wave-number space is, by definition, 
\bea
\nonumber
\<u^i(t,\vk)u^j(t',\vk')\> &=&
\delta(t-t')(2\pi)^d\delta(\vk+\vk')\kappa^{ij}(\vk),\\ 
\label{xi_correlator_k}
\kappa^{ij}(\vk)&=& \kappa(k)\delta^{ij} + \tkappa(k)\,{k_i k_j\over k^2}
\eea
(in the case of incompressible flow, $\tkappa=-\kappa$, 
while for the purely irrotational flow, $\kappa=0$).
We will also sometimes use the spectral energy function 
$I(k) = \kappa^{ii}(k) = d\kappa(k)+\tkappa(k)$.
The magnetic correlation function is defined as follows:
\bea
\nonumber
\<B^i(t,\vk)B^j(t,\vk')\> &=& (2\pi)^d \delta(\vk+\vk')H^{ij}(t,\vk),\\
H^{ij}(t,\vk) &=& {1\over d-1}\,H(t,k)\(\delta^{ij} - {k_i k_j\over k^2}\).
\label{def_Hij}
\eea
The induction equation~\exref{induction_eq} 
in the Fourier space can be written as follows 
(from here on, we suppress the time dependence in the arguments 
unless it is essential for the clarity of exposition):
\bea
\label{ind_eq_k}
\dt B^i(\vk) + \eta k^2 B^i(\vk) =
ik_l\intkp\bigl[u^i(\vk') B^l(\vk-\vk') 
- u^l(\vk') B^i(\vk-\vk')\bigr].
\eea
We can now write an equation for the tensor $B^i(t,\vk)B^j(t,\vk')$ 
and average it: 
\bea
\nonumber
\dt\<B^i(\vk)B^j(\vk')\> 
\,+\, \eta\(k^2+k^{\prime2}\)\<B^i(\vk)B^j(\vk')\> =
\qquad\qquad\\
\nonumber
\intkpp\biggl\{ik_l\Bigl[\<u^i(\vk'') B^l(\vk-\vk'') B^j(\vk')\> 
- \<u^l(\vk'') B^i(\vk-\vk'') B^j(\vk')\>\Bigr]\biggr.
\,\,\,\,\\
\biggl.+\,\,ik'_l\Bigl[\<u^j(\vk'') B^l(\vk'-\vk'') B^i(\vk)\> 
- \<u^l(\vk'') B^j(\vk'-\vk'') B^i(\vk)\>\Bigr]\biggr\} 
\eea
The mixed triple averages that have arisen in the right-hand side 
are split by the Furutsu--Novikov (or ``Gaussian-integration'') 
formula (see \apref{ap_Novikov})  
into products of velocity correlation functions 
defined in the formula~\exref{xi_correlator_k} and averaged response 
functions:
\bea
\nonumber
\<B^i(t,\vk_1) B^j(t,\vk_2) u^l(t,\vk_3)\> &=& 
\int^{t}\diff t'\int\diff^d k' \<u^l(t,\vk_3)u^n(t',\vk')\> 
\<{\delta\[B^i(t,\vk_1)B^j(t,\vk_2)\]\over\delta u^n(t',\vk')}\>\\
&=& {1\over2}\,(2\pi)^d\kappa^{ln}(\vk_3) 
\<{\delta\[B^i(t,\vk_1)B^j(t,\vk_2)\]\over\delta u^n(t,-\vk_3)}\>. 
\eea
The averaged {\em same-time} response function that appears above is
\bea
\nonumber
\<{\delta\[B^i(\vk_1)B^j(\vk_2)\]\over\delta u^n(-\vk_3)}\> =
ik_{1m}\Bigl[\delta^{in}H^{mj}(\vk_1+\vk_3)
- \delta^{mn}H^{ij}(\vk_1+\vk_3)\Bigr]\delta(\vk_1+\vk_2+\vk_3)\\
+\,\,ik_{2m}\Bigl[\delta^{jn}H^{mi}(\vk_2+\vk_3)
- \delta^{mn}H^{ij}(\vk_2+\vk_3)\Bigr]\delta(\vk_1+\vk_2+\vk_3),
\qquad
\label{response_k}
\eea
where the tensor~$H^{ij}(\vk)$ is defined by the formula~\exref{def_Hij}.
The expression~\exref{response_k} is obtained by formally integrating the 
(unaveraged) equation for $B^i(t,\vk_1)B^j(t,\vk_2)$ up to time~$t$, 
taking the functional 
derivative $\delta/\delta u^n(t',\vk_3)$, averaging, setting~$t'=t$, 
and using the causal property of the response functionals.

Using the definitions~\exref{xi_correlator_k} and~\exref{def_Hij}
and after some straightforward algebra, we arrive at the so called 
{\em mode-coupling equation:}
\bea
\nonumber
\dt H + 2(\eta_T+\eta)k^2 H =
\qquad\qquad\qquad\qquad\\
\nonumber
{1\over d-1}\intkpp\Biggl\{(d-1)k^2\kappa(k'')
+ {2(\vk\cdot\vk')(\vk\cdot\vk'')(\vk'\cdot\vk'')\over 
{k'}^2 {k''}^2}\,\tkappa(k'')\\
\nonumber
+ \[k^2 - {(\vk\cdot\vk')^2\over {k'}^2}\]
\bigl[\kappa(k'') + \tkappa(k'')\bigr]
\qquad\qquad\qquad\\
\label{MC_eq}
+\,\,(d-3)\[k^2 - {(\vk\cdot\vk')^2\over {k'}^2}\]\kappa(k'')
+ (d-3)\,{(\vk\cdot\vk'')^2\over {k''}^2}\,\tkappa(k'')\Biggr\}H(k'),
\eea
where $\vk'=\vk-\vk''$, 
and we have introduced an auxiliary quantity that one may term 
{\em the turbulent diffusivity:} 
\bea
\eta_T = {1\over d}\intk I(k)
\eea
[recall that $I(k) = d\kappa(k) + \tkappa(k)$].
\eqref{MC_eq} is the generalization of the mode-coupling equations
of \citet{Kazantsev}, \citet{Vainshtein_SSF-b}, and of \citet{KA} 
to the case of $d$~dimensions 
and arbitrary degree of compressibility. It is now straightforward 
to verify that \eqref{MC_eq} integrated over all wave numbers gives 
the following evolution law of the total magnetic energy 
$W = (2\pi)^{-d}\int{\rm d}^d k\,H(k)$: 
\bea
\label{eq_W}
\dt W = 2\gamma W - 2\eta\,\msk\,W,
\eea
where the growth rate is
\bea
\label{def_gamma}
2\gamma = {d-1\over d}\intk k^2\bigl[2\kappa(k) + \tkappa(k)\bigr]
\eea
and the mean square wave number of the magnetic spectrum is 
\bea
\label{def_msk}
\msk = {1\over W}\,\intk k^2 H(k).
\eea
The non-conservation of the magnetic energy reflected 
by~\eqref{eq_W} is understandable, because the system 
is not closed and the passive magnetic field constantly 
receives energy from the velocity field. 
Note that the white-in-time velocity field formally constitutes an 
infinite reservoir of energy.
   
It is instructive to have also an evolution equation for~$\msk$. 
This is derived in quite the same fashion as \eqref{eq_W} and reads
\bea
\label{eq_msk}
\dt\msk = D + 2\gamma_2\msk - 2\eta\[\overline{k^4} - (\msk)^2\],
\eea
where the growth rate is 
\bea
\label{def_gamma2}
2\gamma_2 = {1\over d}\intk k^2 \[I(k) + 
4\(\kappa(k) + {d+1\over d+2}\,\tkappa(k)\)\],
\eea
the source term, which corresponds to a slow diffusion of the 
magnetic spectrum in $k$~space,~is  
\bea
\label{def_D}
D = {d-1\over d}\intk k^4\bigl[2\kappa(k) + \tkappa(k)\bigr],
\eea
and the ``fourth moment'' of the magnetic spectrum is defined 
analogously to the second~\exref{def_msk}:
\bea
\overline{k^4} = {1\over W}\,\intk k^4 H(k).
\eea
All of the velocity-spectrum moments introduced above 
[formulas~\exref{def_gamma}, 
\exref{def_gamma2}, and~\exref{def_D}] can be shown to be positive 
based on the realizability properties of the isotropic velocity-field 
correlation functions (see \apref{ap_realizability}).  
Note that, since $\overline{k^4} - (\msk)^2 = \overline{(k^2-\msk)^2}\ge0$, 
the resistive term still plays the ultimately balancing role 
in~\eqref{eq_msk}.   

\eqref{eq_W} and \eqref{eq_msk} provide us with a basic understanding  
of the  time-evolution properties of the magnetic energy spectrum. 
Let us assume that initially the magnetic fluctuations are 
concentrated at wave numbers comparable with or greater than 
the characteristic wave number~$\kd$ 
of the velocity field, and that the Prandtl number is large, i.e., 
the characteristic wave number associated with Ohmic resistive diffusion,
$\kres \sim \Pr^{1/2}\kd$, greatly exceeds $\kd$. 
We see that the mean square wave number of the magnetic fluctuations
increases exponentially in time with the growth rate $\gamma_2$, 
which is of the same order as that of the magnetic energy,
until this increase is 
checked by the resistive dissipation [\eqref{eq_msk}]. 
Thus, 
the bulk of the magnetic spectrum will shift towards the resistive 
scale range exponentially fast with characteristic rate $\gamma_2$. 
The fluctuations themselves will also increase in strength, 
exponentially fast as well, with the rate $\gamma$ [\eqref{eq_W}].

Naturally, the question arises, what is going to happen when 
the resistive scales are finally reached. 
That is where the principal 
technical difficulties of this theory lie. 
Before we attack this problem, 
let us probe a little further into the additional information 
\eqref{MC_eq} can yield regarding the properties of the spectra 
of small-scale fields.

\subsection{The Small-Scale-Field Equation}
\label{small_scale_fields}

We saw in the previous section that the spectrum of magnetic fluctuations 
had a tendency to shift to scales much smaller than those of the velocity 
field. This allows us to study the magnetic energy spectra in what we 
shall call {\em the small-scale approximation}. Namely, we may notice 
that the kernel of the integral operator in the right-hand side of 
the mode-coupling equation~\exref{MC_eq} essentially cuts off 
the range of values that $k''$ may take at the scales $k''\sim\kd$, 
and hence, for $k\gg\kd$, we may expand 
\bea
\nonumber
H(k') = H(k) 
&+& \[{1\over2}\({k''\over k}\)^2\sin^2\theta\ 
- {k''\over k}\,\cos\theta\]k H'(k)\\ 
&+& {1\over2}\({k''\over k}\)^2 \cos^2\theta\,k^2 H''(k) 
+ {\cal O}\[(k''/k)^3\],
\eea
where $\theta$~is the angle between~$\vk''$ and~$\vk$.
Expanding similarly the kernel and performing the necessary angle 
integrations (with the aid of~\apref{ap_angle_int}), we arrive at 
the following {\em small-scale-field (SSF) equation:}
\bea
\label{SSF_eq}
\dt H + 2\eta k^2 H = \gamma\(A k^2 H'' + B k H' + C H\),
\eea
where $\gamma$~is defined by the formula~\exref{def_gamma}, while
the dimensionless coefficients $A$, $B$, and~$C$ 
depend only on the dimension of space 
and the degree of compressibility of the velocity field:
\bea
\label{def_A}
A &=& {1+2a\over(d-1)(da+2)},\\
B &=& {d-1 + 4a\over(d-1)(da+2)},\\
C &=& {2(d-2)\over(d-1)(da+2)}.
\eea
Here $a$~is the so-called compressibility parameter of the velocity 
field which arises naturally the small-scale expansion of the velocity 
correlation function and is ubiquitous in one-point statistical theories 
\citep[see \apref{ap_1to2pt} and][]{BS_metric,SK_tcorr}:
\bea
a = \[1+(d+2)\l.\intk k^2\kappa(k)\r/\intk k^2\tkappa(k)\]^{-1}.
\eea
By using the realizability properties of the velocity correlation 
functions (see \apref{ap_realizability}), we find that 
the values of~$a$ are sandwiched between $-1/(d+1)$ (incompressible flow) 
and unity (irrotational flow), $A>0$, the sign of $B$ is undetermined, 
and $C>0$ except in~2D, where it vanishes. 
The values of~$A$, $B$, and~$C$ in the four special cases 
that are usually studied in the literature are given 
in \tabref{SSF_params}.
In the case of incompressible flow, 
\eqref{SSF_eq} was earlier obtained in 3D by \citet{Kazantsev} 
\citep[see also][]{Vainshtein_SSF-a,Vainshtein_SSF-c,KA}
and in $d$~dimensions by \citet{Gruzinov_Cowley_Sudan}.

It is worthwhile noticing that $A(d+1)d + B d + C = 2$, and therefore 
the SSF~equation~\exref{SSF_eq} satisfies the exact energy evolution 
law~\exref{eq_W}. Thus, the small-scale approximation is conservative. 
This is an essential property because it means that 
all of the magnetic energy is concentrated at the small scales. 
It is therefore reasonable to expect that the spectral properties of 
the small-scale fields would mostly be captured by 
the SSF~equation~\exref{SSF_eq}. This consideration encourages 
us to embark on a detailed investigation of the solutions 
of this equation.

\subsection{Spreading of Magnetic Fluctuations in the Wave-Num\-ber Space: 
Diffusion-Free Regime}
\label{spreading}

First, let us pick a wave number~$k_0$ such that $\kd \ll k_0 \ll \kres$, 
and inquire how an initial spectrum concentrated 
at that wave number will evolve over time. This is easily determined, 
because neglecting the diffusive term in~\eqref{SSF_eq} and changing 
to the logarithmic variable~$z=\ln k$, renders this equation 
a one-dimensional heat equation in a moving frame. The Green's function 
for such an equation is well known, so we find that an initial 
spectrum such that $H_0(k)\propto\delta(k-k_0)$ spreads out over time 
into a widening lognormal profile:
\bea
\nonumber
M(t,k) &=& {S_d\over(2\pi)^d}\,k^{d-1}H(t,k)\\ 
\label{M_lognorm}
&=& {W_0\over k_0}\,e^{\lambda_0\gamma t}\({k\over k_0}\)^{\xi_0} 
{1\over\sqrt{4\pi A\gamma t}}\,
\exp\l\{-{\bigl[\ln(k/k_0)\bigr]^2\over4A\gamma t}\r\},
\eea
where $S_d=2\pi^{d/2}/\Gamma(d/2)$ is the area of the unit sphere 
in $d$~dimensions, 
$W_0$~is the initial magnetic energy, and we have introduced 
two new parameters: 
\bea
\label{def_xi0}
\xi_0 &=& d-1 + s_0,
\where s_0 = {A-B\over2A} = {2-d-2a\over2(1+2a)},\\
\label{def_lambda0}
\lambda_0 &=& C  - A s_0^2 = {(3-d)(d-1+4a) + 4(d-2)(1+2a) - (1+2a)^2\over
4(d-1)(1+2a)(da+2)}\\
\nonumber
&=& 2 - A(\xi_0+1)^2.
\eea
The values of~$s_0$, $\xi_0$, and~$\lambda_0$ in the standard cases 
are given in \tabref{SSF_params}.

\placetable{SSF_params}

Examining the solution~\exref{M_lognorm}, we immediately conclude that 
the following basic processes are initiated: (i) the fluctuation strength 
in each logarithmic wave-number band grows or decays 
exponentially at the rate~$\lambda_0\gamma$, 
(ii) the number of such bands that are 
excited [i.e., the width of the lognormal envelope in the 
solution~\exref{M_lognorm}] 
grows exponentially fast at the rate $4A\gamma$, (iii) 
a magnetic spectrum with exponent~$\xi_0$ is formed inside the 
lognormal envelope, with the peak of the spectrum moving rightwards 
(to smaller scales/larger~$k$): $k_{\rm peak} = k_0\exp\(2\xi_0A\gamma t\)$. 
It is not hard to verify that the effective sum of these processes produces 
the total magnetic-energy growth rate we have previously obtained 
[\eqref{eq_W}]: $2\gamma = \lambda_0\gamma + A(\xi_0+1)^2\gamma$. 

\begin{figure}
\plotone{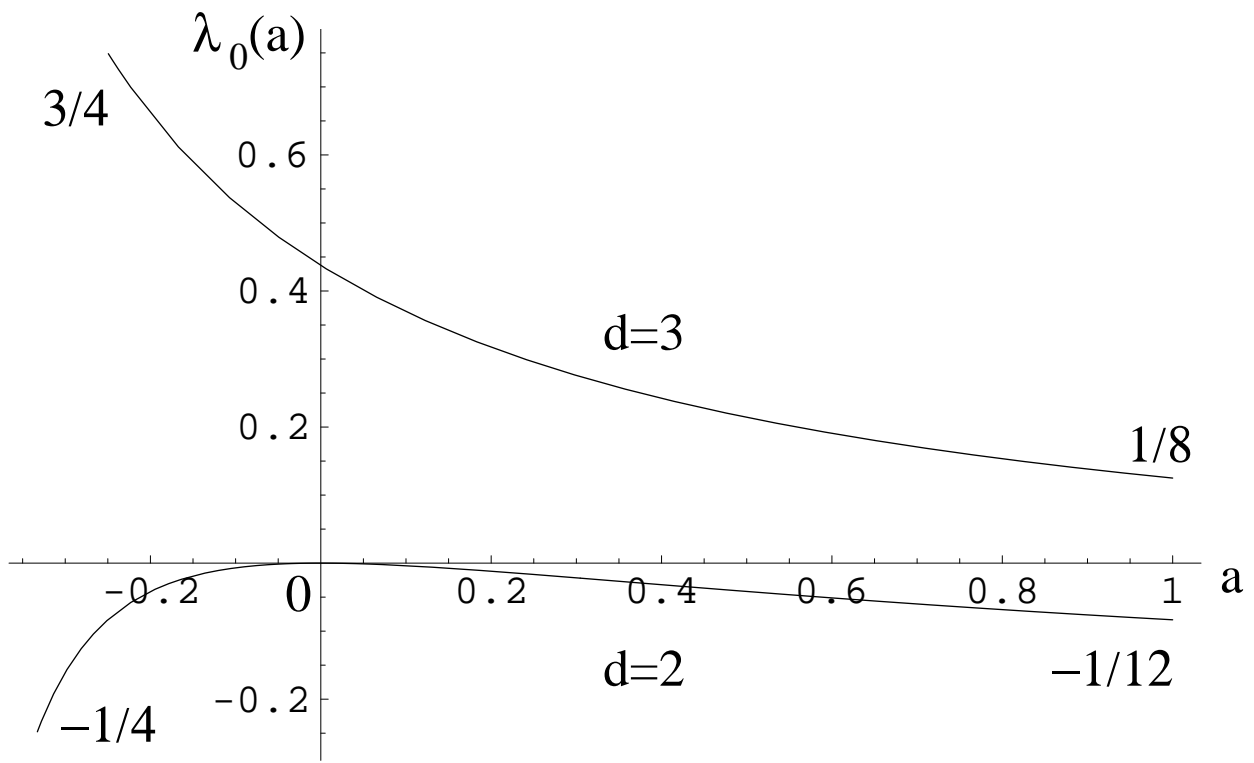}
\caption{Dependence of 
the growth rate of each individual mode,~$\lambda_0$, 
on the compressibility parameter~$a$ in~2 and~3 dimensions 
[see formula~\exref{def_lambda0}] and \tabref{SSF_params}.
\label{fig_lambda0}}
\end{figure}

Let us make several observations regarding the behavior of the 
parameters of the solution~\exref{M_lognorm} defined by 
formulas~\exref{def_xi0} and~\exref{def_lambda0}: the spectral 
exponent~$s_0$ (or~$\xi_0$), and the fraction~$\lambda_0$ of 
the total growth rate that corresponds to the growth or decay 
of each individual Fourier mode. (The latter is plotted 
in~\figref{fig_lambda0} as a function 
of the compressibility parameter~$a$ in two and three dimensions.) 

\placefigure{fig_lambda0}

\noindent (i) In~3D, the spectral slope that forms behind 
the advancing peak of the spectrum does not depend on the 
degree of compressibility: $\xi_0=3/2$ ($s_0=-1/2$) regardless 
of the value of~$a$.   

\noindent (ii) While the number of excited modes increases 
exponentially fast regardless of the particular parameter values, 
individual modes grow in~3D ($\lambda_0>0$ for all~$a$) and 
decay in~2D ($\lambda_0<0$ for all~$a\ne0$).

\noindent (iii) In~2D and for~$a=0$, $\lambda_0=0$ and $s_0=0$. 
Curiously, this suggests the forming of an equilibrium-like, 
equipartion spectrum ($\xi_0=d-1$).

\subsection{Spreading of Magnetic Fluctuations in the Wave-Num\-ber Space: 
Resistive Regime}
\label{spreading_diss}

The solution~\exref{M_lognorm} will cease to be valid when 
the magnetic excitation that has resulted from it 
reaches the integral scales~$\sim\kd$ and/or the resistive 
scales~$\sim\kres$. 
Let us first consider the simpler of these 
two aspects of the problem, namely that where the resistive 
scales are reached~{\em before} the integral ones.
In this case, the further behavior 
of the magnetic fluctuations will be determined by 
the SSF~equation~\exref{SSF_eq} with the resistive term retained.

\begin{figure}
\plotone{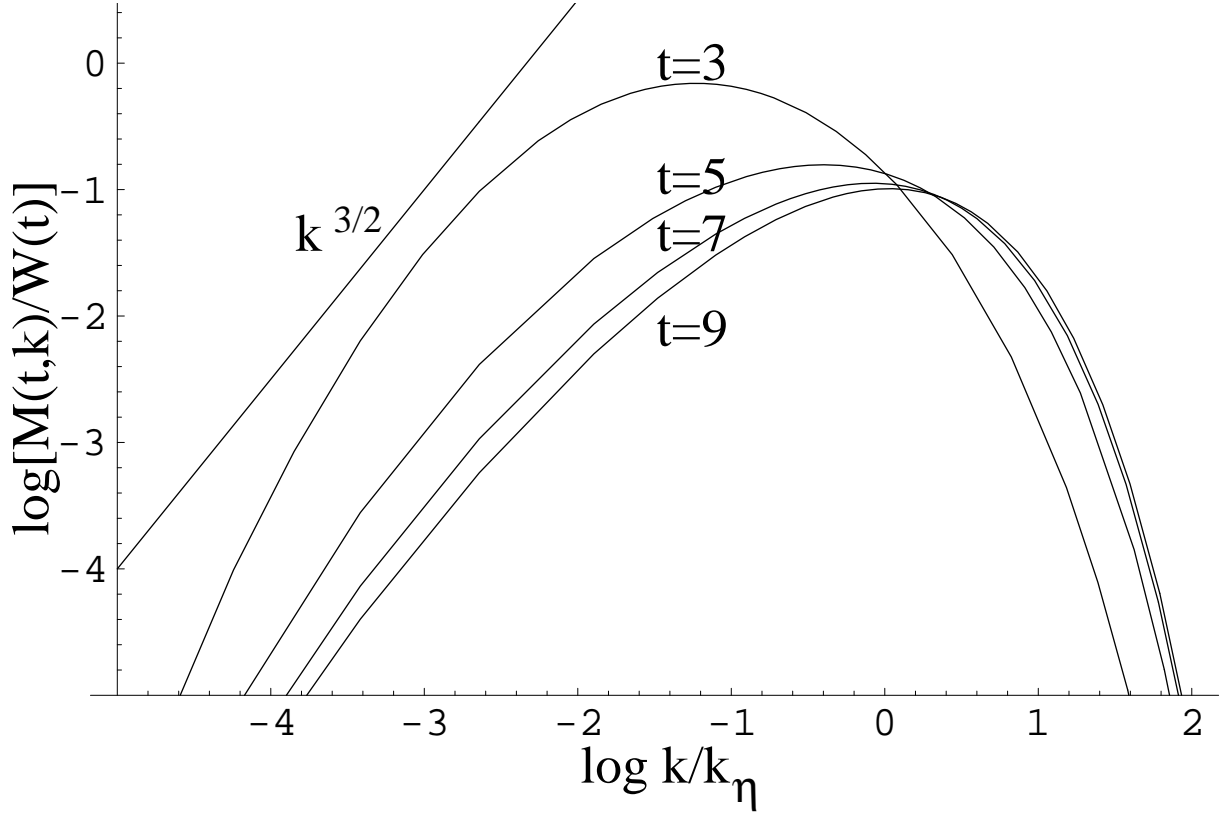}
\caption{The log-log plot of the spectrum~\exref{M_Macdonald} 
in 3D incompressible flow. The spectrum $M(t,k)$ is normalized 
to the total energy~$W(t)$ and shown at times~$3\gamma^{-1}$, 
$5\gamma^{-1}$, $7\gamma^{-1}$, and~$9\gamma^{-1}$. For reference, 
the line corresponding to the expected~$k^{3/2}$ inertial-range 
spectrum is also plotted. The initial wave number was taken 
to be~$k_0=0.05\kres$. All wave numbers are normalized to the resistive 
wave number~$\kres$.
\label{fig_M3log}}
\end{figure}

When the diffusive term is allowed to reappear in the 
SSF~equation~\exref{SSF_eq}, a Green's function solution can still 
be obtained in quadratures. In order to accomplish that, 
we seek the solution in the form
\bea
H(t,k) = e^{\lambda_0\gamma t} k^{s_0} f(t,k/\kres),
\eea
where $\lambda_0$ and $s_0$ are defined by the formulas~\exref{def_lambda0} 
and~\exref{def_xi0}, and $\kres = (\gamma A/2\eta)^{1/2}$~is 
the characteristic wave number of the resistive dissipation. 
It is then elementary to see that $f(t,x)$~satisfies
\bea
\label{f_eq}
\dt f = A\gamma\(x^2 f'' + x f' - x^2 f\).
\eea
The Green's function solution of this problem can be found 
by applying to~\eqref{f_eq} the Kontorovich-Lebedev 
transform \citep[see, e.g.,][]{Erdelyi}, which, in the form adapted 
to our problem, can be defined as follows:
\bea
\label{KL_direct}
F(\tau) &=& \int_0^\infty{\diff x\over x}\,K_{i\tau}(x) f(x)
\quad {\rm (direct),}\\
f(x) &=& {2\over\pi^2}\int_0^\infty\diff\tau\,\tau\sinh(\pi\tau)
K_{i\tau}(x) F(\tau) 
\quad {\rm (inverse),}
\eea
where $K_{i\tau}(x)$~is the modified Bessel function of the second kind 
(also known as the Macdonald, or Basset, function).
Kontorovich-Lebedev transforming~\eqref{f_eq}, we get
\bea
\dt F(t,\tau) = -A\gamma\,\tau^2 F(t,\tau).
\eea
The solution of~\eqref{f_eq} is therefore 
\bea
f(t,x) = {2\over\pi^2}\int_0^\infty{\diff x'\over x'}\,f_0(x') 
\int_0^\infty\diff\tau\,\tau\sinh(\pi\tau)
K_{i\tau}(x)K_{i\tau}(x') e^{-A\gamma\tau^2 t},
\eea
where $f_0(x)$~is the initial profile. 
To compare this result with the inertial-range solution~\exref{M_lognorm}, 
let us recast it in the form of an evolution law for a spectrum 
that is initially a $\delta$-like spike, $M_0(t,k)\propto\delta(k-k_0)$. 
We have
\bea
\label{M_Macdonald}
M(t,k) = {W_0\over k_0}\,e^{\lambda_0\gamma t}\({k\over k_0}\)^{\xi_0} 
{2\over\pi^2}\int_0^\infty\diff\tau\,\tau\sinh(\pi\tau)
K_{i\tau}\({k\over\kres}\)K_{i\tau}\({k_0\over\kres}\) 
e^{-A\gamma\tau^2 t}.
\eea
It is not hard to verify that, for~$k_0\ll\kres$ and~$k\ll\kres$, 
the solution~\exref{M_lognorm} is recovered 
as an asymptotic of the more general solution~\exref{M_Macdonald}. 
On the opposite end of the spectrum, $k\gg\kres$, one finds 
the expected exponential cutoff due to resistive dissipation.
\figref{fig_M3log} illustrates the evolution of 
the spectrum~\exref{M_Macdonald} with time.

\placefigure{fig_M3log}

\begin{figure}
\plotone{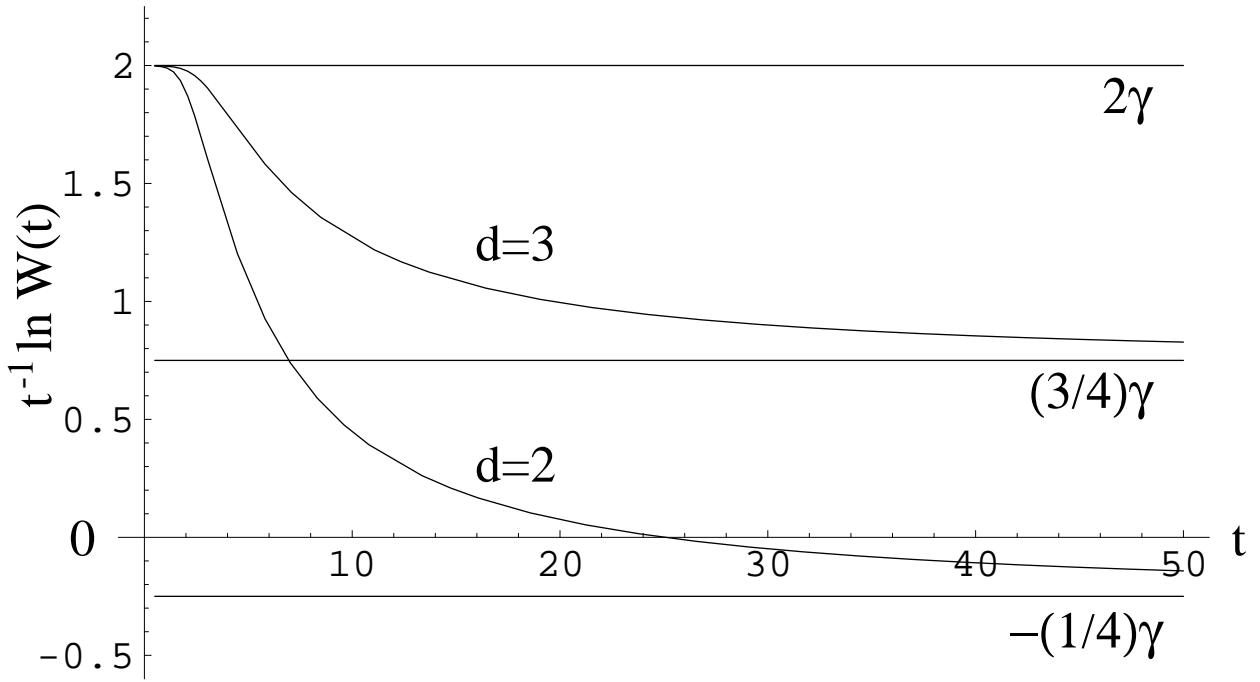}
\caption{Evolution of the ``effective growth rate''  
$\gamma_{\rm eff}(t) = t^{-1}\ln W(t)$ 
of the magnetic energy [see~\eqref{W_evolve}] 
in two and three dimensions in the case of incompressible flow. 
The time is measured in units of~$\gamma^{-1}$ [defined 
by the formula~\exref{def_gamma}].
\label{fig_W23}}
\end{figure}

Integrating the spectrum~\exref{M_Macdonald}, we find that 
the total magnetic energy evolves according~to 
\bea
\nonumber
W(t) &=& W_0\,e^{\lambda_0\gamma t}\(k_0\over\kres\)^{-(\xi_0+1)}\times\\ 
&& \times\,{2^{\xi_0}\over\pi^2}\int_0^\infty\diff\tau\,\tau\sinh(\pi\tau)
\l|\Gamma\({\xi_0+1+i\tau\over2}\)\r|^2K_{i\tau}\({k_0\over\kres}\) 
e^{-A\gamma\tau^2 t}.
\label{W_evolve}
\eea
This formula reflects the gradual slide of the ``effective growth 
rate'' of the magnetic energy, 
$\gamma_{\rm eff}(t) = t^{-1}\ln\bigl[W(t)/W_0\bigr]$, 
from the diffusion-free limit~$\gamma_{\rm eff}(t\to0) = 2\gamma$ 
to the diffusion-reduced value, 
$\gamma_{\rm eff}(t\to\infty) = \lambda_0\gamma$. 
This is demonstrated in \figref{fig_W23}.

\placefigure{fig_W23}

\begin{figure}
\plotone{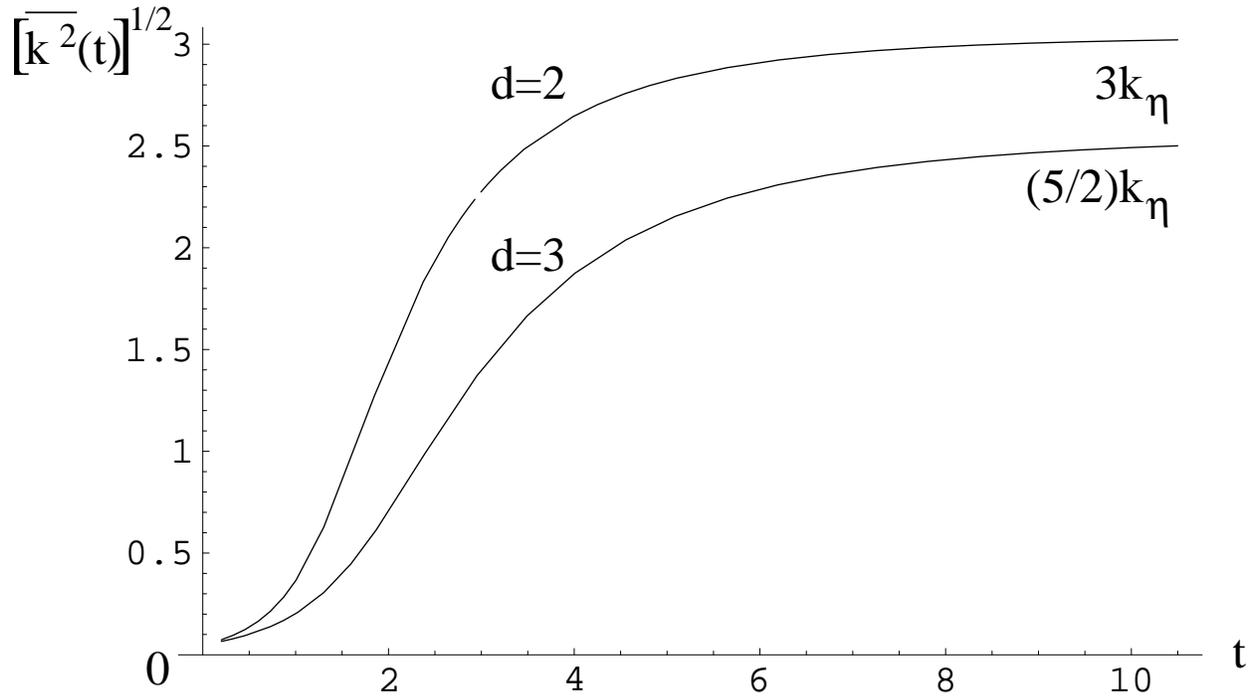}
\caption{Evolution of the mean square wave number 
of the magnetic fluctuations, $\bigl[\msk(t)\bigr]^{1/2}$, 
in two and three dimensions in the case of incompressible flow. 
The time is measured in units of~$\gamma^{-1}$ [defined 
by the formula~\exref{def_gamma}]. The wave numbers are measured 
in units of the inverse resistive scale~$\kres$. 
The initial wave number was taken to be~$k_0=0.05\kres$.
\label{fig_k23}}
\end{figure}

Direct integration of the solution~\exref{M_Macdonald} also 
produces an exact formula for~$\msk(t)$. Its asymptotic behavior 
can, however, be inferred already from that of the energy: 
\bea
\msk(t\to\infty) = {2-\lambda_0\over A}\,\kres^2 
= (\xi_0+1)^2\kres^2
\eea 
[see formula~\exref{def_lambda0}].
This is a good quantitative measure of the characteristic wave number 
in the resistive range around which the bulk of the magnetic 
fluctuation energy will stabilize. The evolution 
of~$\bigl[\msk(t)\bigr]^{1/2}$ is plotted in~\figref{fig_k23}.

\placefigure{fig_k23}

Thus, the solution~\exref{M_Macdonald} describes how the spreading lognormal 
spectrum hits the resistive scale range and and how any further 
refinement of the magnetic-fluctuation scales is suppressed 
by resistive dissipation. The total magnetic energy 
now concentrates in the resistive scale range and 
grows (in~3D) or decays (in~2D) at the rate~$\lambda_0\gamma$.
This regime persists until the excitation finally reaches 
the integral scales. 

\subsection{Spectra and Growth Rates of Magnetic Fluctuations 
in the Long-Time Limit}
\label{quasist_k}

In order to gain some idea of the long-time behavior of the magnetic 
spectra, which sets in when all scales available in the system, 
including the integral ones, are excited, 
let us seek the solution of~\eqref{SSF_eq} in the form 
\bea
\label{quasist_ansatz}
H(t,k) = e^{\lambda\gamma t}h(k).
\eea
The constant spectral profile $h(k)$~must then satisfy the following 
linear ordinary differential equation of the Bessel kind:
\bea
\label{SSF_ev_prob}
Ak^2 h'' + Bk h' + \(C- \lambda - {2\eta\over\gamma}\,k^2\) h = 0.
\eea
Before considering the explicit solutions of this equation, 
it is instructive to recast it in a form mathematically equivalent to that 
of a Schr\"odinger equation for a particle with energy equal to~$-1/\kres^2$
moving in a one-dimensional inverse-square potential,~viz., 
\bea
\label{Schroed_k}
-\psi'' -\[{1\over4} - {\lambda-\lambda_0\over A}\]{1\over k^2}\,\psi 
= -{1\over\kres^2}\,\psi,
\eea
where the ``wave function'' is 
\bea
\psi(k) = k^{-s_0+1/2} h(k),
\eea
$s_0$ and $\lambda_0$~are defined 
by the formulas~\exref{def_xi0} and~\exref{def_lambda0}, 
and $\kres = (\gamma A/2\eta)^{1/2}$ is the inverse resistive scale.
The potential in~\eqref{Schroed_k} 
is repelling and therefore cannot have a negative 
energy level for $\lambda \ge \lambda_0 + A/4$. 
Solutions of the form~\exref{quasist_ansatz} do not exist in this case. 
For $\lambda < \lambda_0$, the so-called ``fall on center'' 
occurs \citep[see, e.g.,][]{Landaushitz3}, 
whereby the solution becomes oscillatory with infinitely many nodes 
as $k$~approaches zero, so it cannot be a spectral function:
for~$k\ll\kres$, 
\bea
\label{subres_as_osc}
\psi \propto \({k\over\kres}\)^{1/2}
\sin\(\sqrt{|\lambda-\lambda_0|\over A}\,\ln{k\over\kres} + \const\).
\eea
Of course, the small-scale approximation is only valid for 
$k \gg \kd \sim \kres\Pr^{-1/2}$, so for $\lambda$~lying 
below~$\lambda_0$, but within some 
$\delta\lambda\sim{\cal O}(1/\ln^2\Pr)$ of it, 
the solution will not yet have any nodes in the range of wave numbers 
for which \eqref{Schroed_k} is valid.     
Thus, the growth rate of the stationary-profile 
solution~\exref{quasist_ansatz} 
is contained in the interval 
\bea
\label{int_lambda}
\lambda_0 - \delta\lambda < \lambda < \lambda_0 + A/4 = \lambdamax, 
\where \delta\lambda \sim {\cal O}\({1\over\ln^2\Pr}\). 
\eea 

For the 3D incompressible case, the bounds~\exref{int_lambda} 
on the growth rate of a stationary-profile spectrum, as well as 
the solution~\exref{sln_K}, were first 
established by \citet{Kazantsev}. He also pointed to 
the fundamental problem that, {\em since the boundary conditions 
at small~$k$ (integral scales) are unknown, there is no rigorous way 
to fix the value of~$\lambda$ within the framework 
of the small-scale approximation.}

The two fundamental solutions of~\eqref{SSF_ev_prob} are the modified 
Bessel functions $I_{\nu(\lambda)}(k/\kres)$ and 
$K_{\nu(\lambda)}(k/\kres)$ times $k^{s_0}$, where 
$\nu(\lambda) = \sqrt{(\lambda-\lambda_0)/A}$. 
The obvious requirement that the physically acceptable solution 
must decay at large~$k$, leaves us with 
\bea
\label{sln_K} 
M(t,k) = {S_d\over(2\pi)^d}\,k^{d-1}H(t,k) 
= C(\lambda)\,e^{\lambda\gamma t}
\({k\over\kres}\)^{\xi_0}K_{\nu(\lambda)}\({k\over\kres}\),
\\\nonumber
C(\lambda) = {W_0\over\kres}\[2^{\xi_0-1}
\Gamma\({\xi_0+1+\nu(\lambda)\over2}\)
\Gamma\({\xi_0+1-\nu(\lambda)\over2}\)\]^{-1},
\eea
where $K_\nu$~is the Macdonald function.
The ``subresistive'' asymptotic of this solution does not depend 
on~$\lambda$ in any essential way and simply 
reflects the expected exponential cutoff of the magnetic spectrum:
\bea
\label{as_superres}
M(t,k\to\infty) \simeq \sqrt{\pi\over2}\,C(\lambda)\,e^{\lambda\gamma t}
\({k\over\kres}\)^{\xi_0-1/2}\exp\(-{k\over\kres}\).
\eea 
The asymptotics valid for~$k\ll\kres$ are 
\bea
\label{as_subres_plus}
M(t,k) &\simeq& 2^{\nu(\lambda)-1}\Gamma\vlp\nu(\lambda)\vrp
C(\lambda)\,e^{\lambda\gamma t}
\({k\over\kres}\)^{\xi_0-\nu(\lambda)}, \quad \lambda > \lambda_0,\\
\label{as_subres_zero}
M(t,k) &\simeq& C(\lambda)\,e^{\lambda_0\gamma t}
\({k\over\kres}\)^{\xi_0}\ln\({\kres\over k}\), \quad \lambda = \lambda_0,
\eea
and, for $\lambda = \lambda_0 - \delta\lambda$, 
where $0<\delta\lambda \ll 1$, we have 
\bea
\label{as_subres_minus}
M(t,k) \simeq \sqrt{\delta\lambda\over A}\,
C(\lambda)\,e^{\lambda\gamma t}
\({k\over\kres}\)^{\xi_0}
\sin\(\sqrt{\delta\lambda\over A}\,\ln{k\over2\kres}\),
\eea
which is, of course, consistent with~\exref{subres_as_osc}.

Clearly, if the initial distribution of the magnetic fluctuations 
is chosen in such a way that it spreads over to resistive scales before 
it does to the integral ones (i.e., if the initial characteristic 
wave number~$k_0$ in the solution~\exref{M_lognorm} 
of~\ssecref{spreading} lies ``closer'' to~$\kres$ than to~$\kd$), 
there will be an intermediate period between the time the resistive 
scales are reached and the time when the excitation finally arrives 
at~$k\sim\kd$. During this period, the boundary condition at the 
integral scales cannot, of course, affect the spectrum. 
The solution of the form~\exref{quasist_ansatz} consistent 
with the spectra obtained in~\ssecref{spreading} and~\ssecref{spreading_diss} 
[solutions~\exref{M_lognorm} and~\exref{M_Macdonald}] 
will clearly be the one with~$\lambda=\lambda_0$: indeed, 
it can be checked that the solution~\exref{sln_K} 
with~$\lambda=\lambda_0$ is the long-time asymptotic of 
the solution~\exref{M_Macdonald}. 
This is essentially what was demonstrated 
numerically by \citet{KA} 
\citep[see also the recent work of][]{Chertkov_etal_dynamo}. 
Our Green's-function solution~\exref{M_Macdonald} is the mathematical 
expression of this fact. 

We may further argue that,
if we were to mandate that the solution~\exref{sln_K} 
vanish at some~$k_0$ such that $\kd\ll k_0\ll\kres$, then,
by virtue of the asymptotic expression~\exref{as_subres_minus}, 
we would be required to set 
\bea
\label{lambda_k0}
\lambda = \lambda_0 - {A\pi^2\over\bigl[\ln(k_0/2\kres)\bigr]^2} 
\simeq \lambda_0 - {A\pi^2\over\bigl[\ln(\Pr^{1/2})\bigr]^2}.
\eea
Note that the particular value of~$k_0$ is unimportant here, 
because $\ln(k_0/2\kres) = \ln(\kd/\kres) + \ln(k_0/2\kd)$, 
the second term being subdominant.
The above value of~$\lambda$ specifies a particular 
solution of the form~\exref{quasist_ansatz} which exactly 
vanishes at~$k=k_0$. In~\ssecref{spreading} and~\ssecref{spreading_diss}, 
we saw however that any excitation present at~$k>k_0$ at a given time 
must necessarily end up spreading over to the integral scales.
The question remains whether, once this happens, the entire 
spectrum would eventually be affected. It is, however, encouraging 
to notice that $\lambda=\lambda_0$~represents a nontrivial limit 
when the infrared cutoff~$k_0$ is taken to zero.

\section{TWO-POINT CORRELATION FUNCTIONS OF THE MAGNETIC FIELD 
IN THE CONFIGURATION SPACE}
\label{sec_corr_fns}

We would now like to put the solution of the kinematic-dynamo 
problem in a somewhat different perspective 
and develop an adequate theory valid in the long-time limit.
In order to make any progress, we necessarily have 
to renounce the small-scale approximation. While it has furnished 
us with a fair amount of practical understanding of the initial 
behavior of the magnetic fluctuations in the kinematic regime, 
it does not contain sufficient information to uniquely 
fix the growth rates and the spectra eventually attained by 
the magnetic energy spectrum. We must therefore attack the 
kinematic dynamo problem in the form that remains valid at all scales. 
This can be more conveniently accomplished in the configuration 
space, for the $\vx$-space analog (inverse Fourier transform)
of the mode-coupling equation~\exref{MC_eq} is {\em local} 
and can be reduced to the quantum mechanics of a particle 
with variable mass in a one-dimensional potential well.
This approach, first proposed by \citet{Kazantsev}, 
leads to a clearer mathematical and physical picture 
of the properties of the magnetic-field correlation functions 
in the long-time limit. 

Kazantsev's quantum mechanics was studied in the case of 
3D~incompressible flow by several authors in 1980s 
\citep{Ruzmaikin_Sokoloff,Novikov_Ruzmaikin_Sokoloff,Artamonova_Sokoloff,Kleeorin_Ruzmaikin_Sokoloff,Maslova_Ruzmaikin}, mostly for particular 
forms of the velocity correlation function. 
\citet{Novikov_Ruzmaikin_Sokoloff} also set forth several important ideas 
with regard to the 2D~case.
In this Section, we develop the configuration-space two-point 
kinematic dynamo theory for a general, $d$-dimensional 
and arbitrarily compressible advecting flow. In~\ssecref{Kazantsev_equation}, 
we derive an evolution equation for the correlation 
function of the magnetic field. This equation is the configuration-space 
analog of the mode-coupling equation of the previous Section. 
We then proceed to formulate the Kazantsev quantum-mechanical 
form of this equation. 
In~\ssecref{quasist_x}, the three principal 
asymptotic regimes of this model are considered: 
the asymptotic solutions are found in the subresistive, inertial, 
and integral scale ranges. In~\ssecref{matching}, 
these solutions are matched in a systematic way and 
the dynamo growth rates are determined for a particular choice 
of small-~and large-scale regularization of the Kazantsev 
quantum mechanics. The ensuing results have universal 
applicability if the specific form of the regularization 
is unimportant. 
The possible relevance of the large-scale 
structure of the velocity correlations is further 
discussed in~\ssecref{discussion}.

\subsection{The Kazantsev Equation}
\label{Kazantsev_equation}

We start with the induction equation in its standard configuration-space 
form~\exref{induction_eq}. The usual assumptions of homogeneity, 
isotropy, and mirror invariance, as well as the Gaussian 
white-noise nature of the velocity field are made [\eqref{velocity_corr}].  
The velocity correlation tensor in the $\vx$~space can be represented 
as follows:
\bea
\label{xi_correlator_x}
\kappa^{ij}(\vy) = \kappaLL(y)\delta^{ij} 
- \(\kappaLL(y)-{\kappaNN(y)\over d-1}\)\(\delta^{ij} - {y^iy^j\over y^2}\),
\eea
where $\kappaLL(y)$ and $\kappaNN(y)$ are called the longitudinal 
and the transverse correlation functions. The tensor~$\kappa^{ij}(\vx)$ 
defined above is the Fourier transform of the $\vk$-space correlation 
tensor~$\kappa^{ij}(\vk)$ whith which we operated in the preceding 
Section [see~\eqref{xi_correlator_k}]. The exact relationship between 
the functions~$\kappaLL(y)$, $\kappaNN(y)$ and the $\vk$-space correlation 
functions~$\kappa(k)$, $\tkappa(k)$ is worked out in detail 
in \apref{ap_iso_FT}. 

We now proceed to derive an evolution equation for the correlation 
tensor $H^{ij}(t,\vy)$ of the magnetic field 
[see definition~\exref{mfield_corr}] by directly averaging 
the dynamic equation for the tensor~$B^i(t,\vx)B^j(t,\vx')$ 
(here, as always, $\vx-\vx'=\vy$) and splitting the arising triple 
averages 
according to the Furutsu--Novikov formula (see \apref{ap_Novikov}). 
Namely, 
\bea
\label{H_triple}
\dt H^{ij}(\vy) - 2\eta\Delta H^{ij}(\vy) = 
{\d\over\d y^k}\Bigl[C^{ikj}(\vy) - C^{kij}(\vy) 
- C^{jki}(-\vy) + C^{kji}(-\vy)\Bigr], 
\eea
where the four terms on the right-hand side are the triple averages: 
\bea
\nonumber
C^{kij}(\vy) &=& C^{kij}(\vx_1-\vx_2) 
= \<u^k(\vx_1)B^i(\vx_1)B^j(\vx_2)\>\\
&=& {1\over2}\[\vblp\kappa^{kl}(\vy) - \kappa^{kl}(0)\vbrp H^{ij}_{,l}(\vy) 
- \kappa^{kj}_{,l}(\vy) H^{il}(\vy) + \kappa^{kl}_{,l} H^{kj}(\vy)\].
\eea
The details of the derivation of the above expression 
are standard and straightforward and have therefore been omitted.  
The reader is reminded that lower indices occurring after a comma 
designate derivatives with respect to~$\vy$. 
Upon collecting all terms in~\eqref{H_triple}, we arrive at the following 
equation for the correlation tensor~$H^{ij}(t,\vy)$ 
\citep[cf.][]{Vainshtein_SSF-a,Vainshtein_SSF-c,MRS_equations,Kichatinov,Vainshtein_Kichatinov}:
\bea
\nonumber
\dt H^{ij} - 2\eta\Delta H^{ij} &=& 
-\vblp\kappa^{kl}-\kappa^{kl}(0)\vbrp H^{ij}_{,kl} 
- \kappa^{ij}_{,kl} H^{kl} 
+ \kappa^{il}_{,k} H^{kj}_{,l} + \kappa^{jl}_{,k} H^{ki}_{,l}\\
&& - 2\kappa^{kl}_{,k} H^{ij}_{,l}
- \kappa^{kl}_{,kl} H^{ij} 
+ \kappa^{il}_{,kl} H^{kj} + \kappa^{jl}_{,kl} H^{ki}.
\label{eq_Hij}
\eea

Since the magnetic field is isotropic and solenoidal ($H^{ij}_{,j}=0$), 
all the necessary information about its two-point correlation properties 
is contained in just one scalar function of~$y$. The most convenient 
(and customary) choice is the longitudinal correlation 
function 
\bea
\label{def_H_LL}
H_{LL}(t,y) = {y^iy^j\over y^2}\,H^{ij}(t,\vy), 
\eea
in terms of which the correlation 
tensor~$H^{ij}(t,\vy)$ can be expressed as follows 
(see \apref{ap_sol_pot}): 
\bea
\label{Hij_H_LL}
H^{ij}(t,\vy) = H_{LL}(t,y)\delta^{ij} 
+{1\over d-1}\,y H_{LL}'(t,y)\(\delta^{ij} - {y^iy^j\over y^2}\).
\eea
Here and below primes denote derivatives with respect to~$y$.
Multiplying~\eqref{eq_Hij} by $y^iy^j/y^2$ and making use of 
the formulas~\exref{def_H_LL} and~\exref{Hij_H_LL}, 
and after more tedious algebra, we find that 
the longitudinal correlation function of the magnetic field 
satisfies the following evolution equation:
\bea
\nonumber
\dt H_{LL} = K(y) H_{LL}'' 
+ \(2\,{d-1\over y}\,K(y) + K'(y) + {3-d\over y}\,Q(y)\)H_{LL}'
\quad\\ 
+\,\,{d-1\over y}\(K'(y) + Q'(y) + 
{d-2\over y}\,\bigl[K(y) - Q(y)\bigr]\)\,H_{LL},
\label{eq_H_LL}
\eea
where we have introduced two ``renormalized diffusivities''
\bea
K(y) &=& 2\eta + \kappaLL(0) - \kappaLL(y),\\
Q(y) &=& 2\eta + {\kappaNN(0) -\kappaNN(y)\over d-1} 
= 2\eta + \kappaLL(0) - {\kappaNN(y)\over d-1}.
\eea
Note that, in the incompressible case, we should have 
(see \apref{ap_sol_pot})
\bea
\label{QK_sol}
Q(y) = K(y) + {yK'(y)\over d-1}.
\eea
Due to the realizability constraints, it is assured that~$K(y)>0$,
$Q(y)>0$ (see \apref{ap_realizability}).
\eqref{eq_H_LL} is the configuration-space analog of the mode-coupling 
equation~\exref{MC_eq}. While we have chosen to derive it by directly 
averaging the induction equation, it could also have been done by 
Fourier-transforming the mode-coupling 
equation~\exref{MC_eq}. We should like 
to remark, however, that \eqref{eq_H_LL} is more convenient 
than the mode-coupling equation~\exref{MC_eq} for the study of 
general velocity correlators, such as those whose small-scale expansion 
is nonanalytic (as in the case of $\Pr<1$). 

A further modification of~\eqref{eq_H_LL} brings us to the quantum-mechanical
analogy that was mentioned in the Introduction. 
Namely, we introduce the ``$\psi$-function''
\bea
\label{def_psi}
\psi(t,y) = y^{d-1}\sqrt{K(y)}\,H_{LL}(t,y)
\exp\({3-d\over2}\int^y{\rm d}y'{Q(y')\over y'K(y')}\)
\eea
and find that it satisfies the one-dimensional Schr\"odinger equation,
\bea
\label{Kaz_eq}
\dt\psi = {1\over 2m(y)}\,\psi'' - V(y)\psi, \qquad y\ge0,
\eea
which describes the quantum mechanics of a particle with variable 
mass\footnote{The mass~$m(y)$ is always positive due to the realizability 
constraint~\exref{realizability_kappaLL} 
(see~\apref{ap_realizability}).}~$m(y)= \[2K(y)\]^{-1}$ 
in a one-dimensional (or radial) potential 
\bea
\label{potential}
V(y) = {1\over2}\,K''(y) - {\[K'(y)\]^2\over 4K(y)} 
+ {3d-5\over 2y^2}\,\[Q(y) - yQ'(y)\] 
+ {(d-3)^2\over 4y^2}\,{\[Q(y)\]^2\over K(y)}.
\eea
\eqref{Kaz_eq} [with potential~\exref{potential}] was first derived
by \citet{BS_dynamo}.
In the three-dimensional case and for incompressible velocity 
field, it reduces to the equation derived by \citet{Kazantsev} 
\citep[see also][]{Kichatinov,Vainshtein_Kichatinov}.
In what follows, the general equation~\exref{Kaz_eq} will also be referred 
to as {\em the Kazantsev equation.} 
The expression for the potential~\exref{potential} in the incompressible 
case becomes
[see formula~\exref{QK_sol}]
\bea
\label{potential_inc}
V(y) = {d^2-1\over 4y^2}\,K(y) 
-{d-2\over d-1}\[K''(y) + {d+1\over y}\,K'(y)
+ {\[K'(y)\]^2\over (d-1) K(y)}\].
\eea
It is this incompressible form that 
has mostly been considered in the literature \citep{Ruzmaikin_Sokoloff,Novikov_Ruzmaikin_Sokoloff,Artamonova_Sokoloff,Kleeorin_Ruzmaikin_Sokoloff,Maslova_Ruzmaikin}. Note that, in 2D incompressible case, it 
is immediately clear from the formula~\exref{potential_inc} that 
the potential is repelling and hence no bound states exist.

\subsection{The Eigenvalue Problem Associated with the Kinematic Dynamo: 
Asymptotic Solutions}
\label{quasist_x}

We now consider the eigenvalue problem associated with the Kazantsev equation:
\bea
\label{Kaz_ev_prob}
-\lambda\gamma\psi = -{1\over 2m(y)}\,\psi'' + V(y)\psi, \qquad y\ge0,
\eea 
where, for the sake of theoretical uniformity, we measure 
the eigenvalues (dynamo growth rates) in terms of~$\gamma$, 
the magnetic energy growth rate~\exref{def_gamma} that 
figured in \secref{sec_spectra}. [In terms of the configuration-space 
quantities, $\gamma$~is proportional to the square of 
the velocity shear:~$\gamma=|\kappaLL''(0)|/2A$, 
where~$A$ is defined by the formula~\exref{def_A}. See~\apref{ap_1to2pt}, 
formula~\exref{gamma_kA}.] The objective is to find 
the spectrum of the Schr\"odinger operator in the right-hand side 
of~\eqref{Kaz_ev_prob}. Clearly, positive energies correspond to 
damped modes ($\lambda<0$), while negative energies represent 
growth (the dynamo effect, $\lambda>0$). Therefore, the ground state 
of~\eqref{Kaz_ev_prob}, if it exists, would give the desired long-time 
asymptotic of the magnetic-field correlation function.

At small scales, the velocity correlation tensor~\exref{xi_correlator_x} 
has a Taylor expansion,
\bea
\kappa^{ij}(\vy) = \kappa_0\delta^{ij} 
-{1\over 2}\,\kappa_2\(y^2\delta^{ij} + 2ay^iy^j\) + \cdots,
\eea
and the renormalized diffusivities~$K(y)$ and~$Q(y)$ 
therefore are, to second order (see \apref{ap_1to2pt}),
\bea
K(y) &\simeq& 2\eta + {1\over2}\,(1+2a)\kappa_2 y^2,\\
Q(y) &\simeq& 2\eta + {1\over2}\,\kappa_2 y^2,
\eea
where $a$~is the compressibility parameter. 
These are valid for~$y \ll (\kappa_0/\kappa_2)^{1/2} \sim \kd^{-1}$, 
where $\kappa_0 = \kappaLL(0)$. 
Note that $(1+2a)\kappa_2 = |\kappaLL''(0)|= 2\gamma A$ 
[see formula~\exref{gamma_kA}].

When $y \gg (\kappa_0/\kappa_2)^{1/2}$, both~$K(y)$ and~$Q(y)$ tend 
to a constant value:
\bea
K(y) \simeq Q(y) \simeq 2\eta + \kappa_0 \simeq \kappa_0. 
\eea 

\begin{figure} 
\plotone{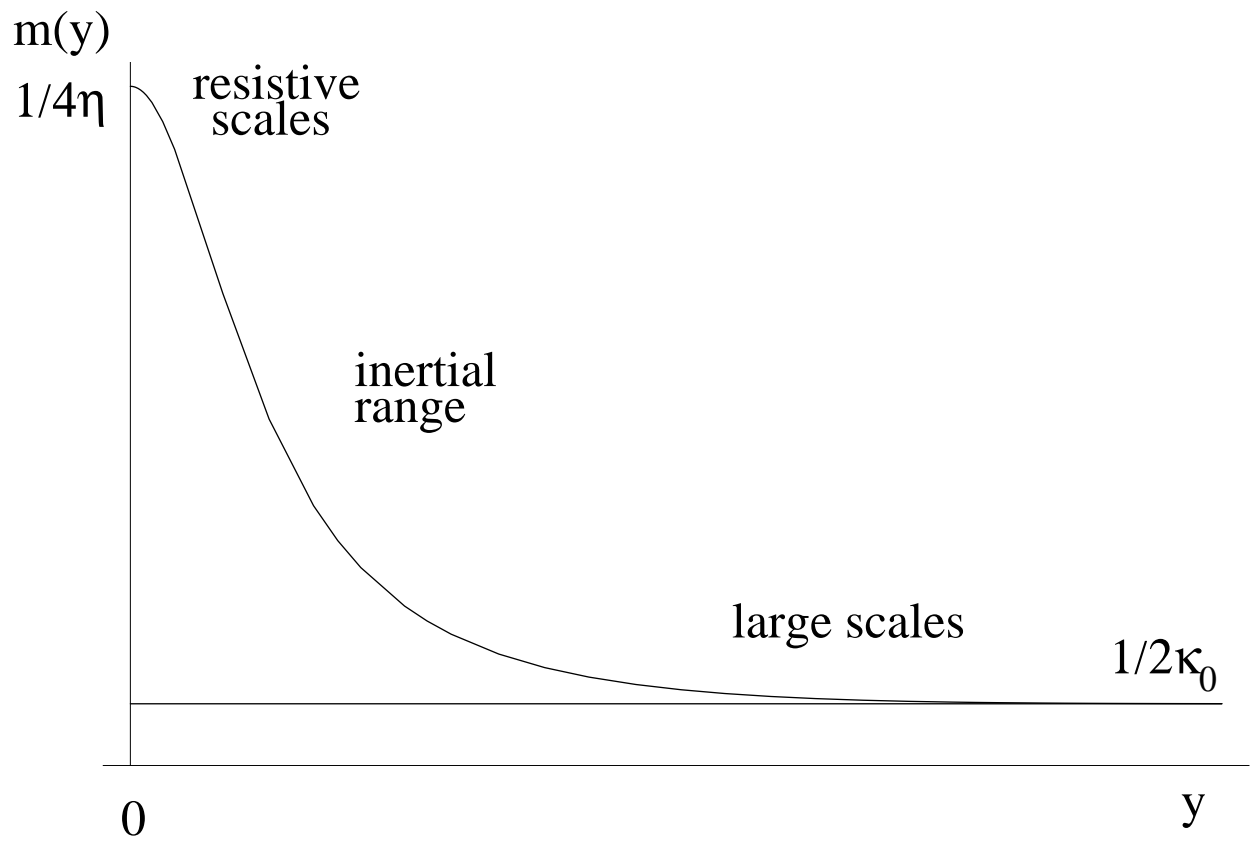}
\caption{A schematic plot of the variable mass in Kazantsev 
quantum mechanics.
\label{fig_mass}}
\end{figure}

\begin{figure}
\plotone{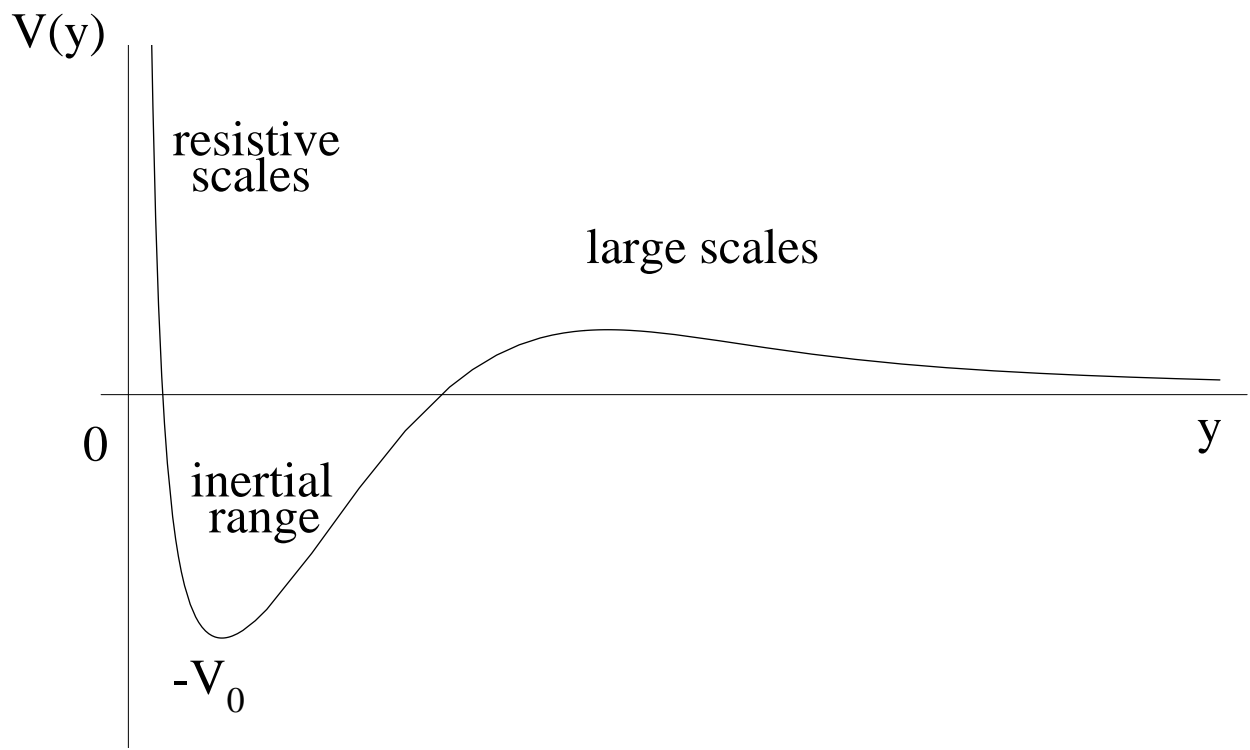}
\caption{A schematic plot of the potential in Kazantsev 
quantum mechanics in three dimensions [see formula~\exref{potential}]. 
In its particulars, this plot depicts the potential~$V(y)$ 
for the~3D~incompressible flow.
\label{fig_potential}}
\end{figure}

Substituting these small- and large-scale asymptotics of~$K(y)$ and~$Q(y)$ 
into the expressions for the mass of our particle and for the potential 
it lives in [see~\exref{potential}], we may gain a qualitative idea 
of what~$m(y)$ and~$V(y)$ look like. 
[The typical forms of the mass and potential (in~3D) are sketched in 
\figref{fig_mass} and \figref{fig_potential}.]
Three regions with different 
asymptotics are clearly pronounced: {\em the subresistive scales,} 
{\em the integral scales,} and the intermediate {\em inertial range.} 
Let us analyze the situation in each of them in turn.


\placefigure{fig_mass}
\placefigure{fig_potential}
\notetoeditor{These two figures should appear side-by-side in print}

\paragraph{The subresistive scales.} These are, of course, 
the scales satisfying $y \ll (\eta/\kappa_2)^{1/2} \sim \kres^{-1}$. 
There, the mass of the particle is constant and the potential 
is a repelling inverse square:
\bea
m(y) &\simeq& {1\over4\eta},\\
V(y) &\simeq& {1\over 2m(0)}\,{d^2-1\over4y^2} \to +\infty, 
\quad y\to +0.
\eea
The asymptotic solutions of the eigenvalue problem~\exref{Kaz_ev_prob}  
in this approximation are
\bea
\label{sln_subres}
\psi(y) = y^{(1\pm d)/2},
\eea
where it is clearly the ``$+$''-branch that must be picked. 
Indeed, it satisfies just the right boundary condition that follows 
from the definition of the ``wave function''~\exref{def_psi} 
and the requirement that $H_{LL}(t,y)$~have a positive finite 
value at~$y=0$ (it is the total magnetic energy in the system).
Note that {\em the boundary condition could not have been satisfied 
had there not been a nonzero magnetic diffusivity in the problem.}

The solution in this region does not depend on the value 
of~$\lambda$, in a precise analogy to what we saw in the wave-number 
space [see the asymptotic~\exref{as_superres}].

\paragraph{The integral scales.} Considering the opposite end of 
the available range of scales, 
for~$y \gg (\kappa_0/\kappa_2)^{1/2} \sim \kd^{-1}$, we 
again find a constant (now ``renormalized'') mass and 
a repelling inverse-square potential:
\bea
m(y) &\simeq& {1\over2(2\eta + \kappa_0)} \simeq {1\over2\kappa_0},\\
V(y) &\simeq& {1\over 2m(\infty)}\,{d^2-1\over4y^2} \to +0, 
\quad y\to +\infty.
\eea
This time, however, it is the potential that contributes subdominantly 
to the solution, while the eigenvalue~$\lambda$ plays an important 
role (unless~$\lambda=0$). The solutions that do not grow at 
infinity are 
\bea
\label{sln_int_plus}
\psi(y) &=& \exp\(-\sqrt{\lambda\gamma\over\kappa_0}\,y\), 
\qquad \lambda>0,\\
\label{sln_int_zero}
\psi(y) &=& y^{-(d-1)/2},
\qquad \lambda=0,\\
\label{sln_int_minus}
\psi(y) &=& \sin\Biggl(\sqrt{|\lambda|\gamma\over\kappa_0}\,y +\phi\Biggr), 
\qquad \lambda<0.
\eea
We see from the form of the solution, as well as from the fact that 
the potential approaches zero from above as~$y\to +\infty$, 
that {\em no bound states can exist with negative~$\lambda$.} 
In fact, {\em neither can they exist with~$\lambda=0$,} for it leads to 
a power-like decay of the wave function at best.

\paragraph{The inertial range.} These are the scales that 
are equally removed both from where resistive-diffusive effects 
operate and from the integral scales where the details of the velocity-field 
correlation properties start to be felt. Namely, 
$(\eta/\kappa_2)^{1/2} \ll y \ll (\kappa_0/\kappa_2)^{1/2}$. 
In this regime, we obtain
\bea
\label{m_inert}
m(y) &\simeq& {1\over(1+2a)\kappa_2 y^2},\\
\label{V_inert}
V(y) &\simeq& - V_0, \where 
V_0 = {\kappa_2\over4}\[3d-5 - {(d-3)^2\over2(1+2a)}\].
\eea
We have clearly found ourselves inside the potential well, 
which is locally flat. An obvious estimate for 
the eigenvalue immediately follows: $\lambda < V_0/\gamma = \lambdamax$, 
which is the same as the upper bound obtained in~\ssecref{quasist_k} 
[see the inequality~\exref{int_lambda}]. We find that, in~$d=3$, 
the potential well always extends below zero (so $V_0>0$), 
while in~$d=2$, $V_0\le0$ for $a \le -1/4$. 

\eqref{Kaz_ev_prob} in this regime takes the form 
\bea
\label{eqn_inert}
\psi'' + {\lambdamax-\lambda\over Ay^2}\,\psi = 0, 
\eea
where~$A$ is defined by the formula~\exref{def_A} and it can be easily 
verified that~$(1+2a)\kappa_2 = 2\gamma A$. The effective potential 
in~\eqref{eqn_inert} 
is of the inverse-square kind, as advertised in the Introduction 
(\ssecref{outline}). This equation 
is, of course, the configuration-space version of~\eqref{Schroed_k} 
in the inertial-scale limit.
Its solutions are, expectedly,
\bea
\label{sln_inert}
\psi(y) = c_1 y^{1/2+\sqrt{(\lambda-\lambda_0)/A}}
+ c_2 y^{1/2-\sqrt{(\lambda-\lambda_0)/A}},
\eea 
where~$\lambda_0 = \lambdamax-A/4$ is defined 
as it was in~\ssecref{quasist_k} [see formula~\exref{def_lambda0}].

\subsection{Asymptotic Matching: 
The Dynamo Growth Rates in the Long-Time Limit}
\label{matching}

In order to complete the asymptotic solution 
of the problem, one must now proceed to properly match the solutions 
we have obtained in the three asymptotic scale ranges described above. 

Matching the inertial range solution with the subresistive one 
does not present any difficulty in principle. 
Indeed, we saw in~\secref{sec_spectra} that, in the $\vk$~space, 
an analytic solution could be obtained that is valid both in the resistive 
and inertial ranges. 
The issue of matching the inertial-range 
and the large-scale solutions is more troublesome. 
The appropriate matching procedure depends on the 
full functional form of the potential at the velocity-field 
scales and must typically be performed assuming a particular 
form of the velocity correlator 
\citep[cf.][]{Artamonova_Sokoloff,Kleeorin_Ruzmaikin_Sokoloff}. 

In this section, we propose a simple way to match the asymptotic 
solutions obtained in~\ssecref{quasist_x} and calculate the 
dynamo growth rates in the long-time limit. Our method is based 
on the assumption that the particular way the Kazantsev particle 
mass and potential are regularized at small and large scales 
does not play an essential part in determining the dynamo growth 
rates. This assumption holds for such forms of the velocity correlation 
function that ensure that the potential well be located 
at small (inertial) scales and that the potential~\exref{potential} 
have no relevant structure at the integral scales.
In other words, our model quantum-mechanical particle 
must be ``localized'' at small scales in order for the 
structure of the velocity correlations to be unimportant.
If this is satisfied, the dynamo growth rates in the long-time 
limit can be calculated in a universal way.
Let us explain how this can be accomplished. 


Although, as we have noted, matching the subresistive 
and the inertial-range solutions is no problem in principle, 
it does require a considerable amount of 
algebra \citep{Kleeorin_Ruzmaikin_Sokoloff}. The formal difficulty 
lies in the fact that $\psi(y)$ and $\psi'(y)$ at 
$y \sim (\eta/\kappa_2)^{1/2}$ are of different orders in~$\eta$. 
The matching must therefore be carried out in two successive 
orders of the asymptotic expansion. This difficulty can 
be circumvented in the following way, suggested by Landau's 
discussion of the quantum mechanics in the inverse-square 
potential \citep{Landaushitz3}. Assuming that the particular 
form of the small-scale regularization is not essential 
for choosing the right solution in the inertial range, 
let us pick a small $\yres \sim (\eta/\kappa_2)^{1/2}$, 
extend the inertial-range forms of the mass and 
potential [formulas~\exref{m_inert} and~\exref{V_inert}] 
down to this point (to all~$y>\yres$), and replace the actual values 
of~$m(y)$ and~$V(y)$ at~$y<\yres$ by constants equal to 
their values at~$y=\yres$ (thus taking ``a shortcut'' to the 
``wall'' at~$y=0$). Let us then solve the resulting 
eigenvalue problem in both regions and match the solutions 
at~$y=\yres$. We will then carry out a similar procedure 
to match the inertial-range and the large-scale solutions 
at some~$y_0 \sim (\kappa_0/\kappa_2)^{1/2}$ (see below). 
Eventually, we must take the limit~$\yres\to0$ and~$y_0\to\infty$ and 
determine the eigenvalues~$\lambda$ from the requirement 
that the matching relations at~$\yres$ and~$y_0$ be consistent 
with each other in this limit. 

Let us empahsize that our matching procedure corresponds 
to a particular reasonable choice of the small-~and large-scale 
regularization of the Kazantsev quantum mechanics. Its 
universal applicapility hinges on the assumption that 
the specific form of such regularization is inessential. 

The proper solution in the interval~$0<y<\yres$ is 
$\psi(y) = \sin(\gres y/\yres)$, where 
$\gres=\sqrt{(\lambdamax-\lambda)/A}$ 
(we assume, of course, that~$\lambda<\lambdamax$). 
The solution at~$y>\yres$ is given by the formula~\exref{sln_inert}.
Matching $\psi(\yres)$ and $\psi'(\yres)$, we get, for~$\lambda>\lambda_0$, 
\bea
\label{match_c1c2}
{c_2\over c_1} = 
{s_1-\gres\cot\gres\over\gres\cot\gres-s_2}\,\yres^{s_1-s_2},
\qquad \lambda_0<\lambda<\lambdamax.
\eea
Here we have denoted~$s_{1,2}=1/2\pm\sqrt{(\lambda-\lambda_0)/A}$.
For~$\lambda<\lambda_0$, it is convenient to write 
the inertial-range solution~\exref{sln_inert}~as 
\bea
\label{sln_inert_sin}
\psi(y) = c\,y^{1/2}
\sin\[\sqrt{\delta\lambda\over A}\,\ln\({y\over\yres}\) + \chi\], 
\qquad \delta\lambda = \lambda_0-\lambda>0.
\eea
The matching determines the value of the phase:
\bea
\label{match_phase}
\cot\chi = {\gres\cot\gres-1/2\over\sqrt{\delta\lambda/A}},  
\where \gres = {1\over2}\,\sqrt{1+{4\delta\lambda\over A}}.
\eea

Let us now employ the same device to match the inertial-range and 
the large-scale solutions. Namely, we now pick 
some~$y_0\sim(\kappa_0/\kappa_2)^{1/2}$, 
formally extend the validity of the inertial-range asymptotics 
of~$m(y)$ and~$V(y)$ to all~$y<y_0$, while at~$y>y_0$,  
let $m(y)=1/2\kappa_0$ and $V(y)=0$. The solution in the latter region 
is the large-scale solution~\exref{sln_int_plus} or~\exref{sln_int_minus}
(depending on the sign of~$\lambda$). For convenience, 
let us take~$y_0=(\kappa_0/\gamma A)^{1/2}$, whence, 
in the formulas~\exref{sln_int_plus} and~\exref{sln_int_minus}, 
we have, by definition, 
$\lambda\gamma/\kappa_0 =\lambda/A y_0^2 = (\gamma_0/y_0)^2$, 
where~$\gamma_0=\sqrt{|\lambda|/A}$. 
In order to match~$\psi(y_0)$ and~$\psi'(y_0)$, we take 
the asymptotic solutions appropriate for various values 
of~$\lambda$: for~$\lambda>\lambda_0$, the formulas~\exref{sln_inert}, 
\exref{sln_int_plus}, and~\exref{sln_int_minus} imply 
\bea
\label{match_gg}
{c_1\over c_2} &=& - {s_2+\gamma_0\over s_1+\gamma_0}\,y_0^{s_2-s_1}, 
\qquad \lambda>\lambda_0~\&~\lambda\ge0,\\
\label{match_gl}
{c_1\over c_2} &=& 
{s_2 - \gamma_0\cot(\gamma_0+\phi)\over\gamma_0\cot(\gamma_0+\phi) -s_1}\,
y_0^{s_2-s_1}, 
\qquad \lambda_0<\lambda<0;
\eea
for~$\lambda<\lambda_0$, we use the formula~\exref{sln_inert_sin} 
instead of~\exref{sln_inert} and get 
\bea
\label{match_lg}
\cot\[\sqrt{\delta\lambda\over A}\,\ln\(y_0\over\yres\) + \chi\] 
&=& -{\gamma_0+1/2\over\sqrt{\delta\lambda/A}}, 
\qquad 0\le\lambda<\lambda_0,\\
\label{match_ll}
\cot\[\sqrt{\delta\lambda\over A}\,\ln\(y_0\over\yres\) + \chi\] 
&=& {\gamma_0\cot(\gamma_0+\phi)-1/2\over\sqrt{\delta\lambda/A}},
\qquad \lambda<\lambda_0~\&~\lambda<0,
\eea 
where~$\delta\lambda=\lambda_0-\lambda$. 
Note that the relations~\exref{match_gg} and~\exref{match_lg} smoothly 
extend to the case of~$\lambda=0$ (where $\gamma_0=0$), which 
does not have to be treated separately.

We are now in a position to find the dynamo growth rates. 
In the matching conditions derived above, we take the limit 
$\yres/y_0\to0$ and determine the allowed values of~$\lambda$ 
from the consistency requirements.

\paragraph{The case $\lambda>\lambda_0~\&~\lambda\ge0$ (growth).} 
For these values of~$\lambda$, the relations~\exref{match_c1c2}  
and~\exref{match_gg} that ensure proper matching of 
the inertial-range solution at dissipative and integral scales
respectively are incompatible in the limit~$\yres/y_0\to0$ unless 
\bea
\label{consist_gg}
(\gamma_0 + s_1)(\gres\cot\gres - s_2) = 0.
\eea
This is a transcendental equation on~$\lambda$, which is 
readily seen to have no solutions.   
Thus, {\em no positive growth rates that are larger 
than~$\lambda_0\gamma$ are allowed.} 

Note that, in~2D, while we have~$\lambda_0\le0$, a rudimentary potential 
well exists for some values of the compressibility 
parameter: in~\ssecref{quasist_x}, we already pointed out 
that~$\lambdamax>0$ for~$a>-1/4$. We see now that this well has 
turned out to be too shallow to sustain a bound state 
with~$\lambda\ge0\ge\lambda_0$.

\paragraph{The case $0\le\lambda<\lambda_0$ (growth).} 
In this case, we have to make sure that 
the relations~\exref{match_phase} and~\exref{match_lg} 
are consistent with each other in 
the limit~$y_0/\yres\sim\Pr^{1/2}\to\infty$. 
If~$\delta\lambda=\lambda-\lambda_0$ remains finite under this limit, 
the left-hand side of~\eqref{match_lg} has no definite limit, 
while its right-hand side is equal to a finite constant. 
The matching is therefore impossible. 
On the other hand, if we assume that~$\delta\lambda\to+0$ 
as~$y_0/\yres\to\infty$, the right-hand sides of both 
relations~\exref{match_lg} and~\exref{match_phase} tend to infinity, 
whence~$\chi=\pi m$,~$m\in{\mathbb Z}$, and  
\bea
\label{deltalambda_n}
\delta\lambda = {A\pi^2n^2\over\bigl[\ln(y_0/\yres)\bigr]^2} 
\simeq {A\pi^2n^2\over\bigl[\ln(\Pr^{1/2})\bigr]^2} \ll 1, 
\qquad n=1,2,\dots 
\eea 
Clearly, it does not matter here 
what the particular definitions of~$y_0$ and~$\yres$ are: 
multiplying any given such definitions by finite constants 
leads to a finite, and hence subdominant, correction in 
the denominator of the expression~\exref{deltalambda_n}. 

The expression~\exref{deltalambda_n} reproduces 
the result that was obtained in~\ssecref{quasist_k} 
[see formula~\exref{lambda_k0}] by requiring 
the magnetic-energy spectrum to vanish at some 
particular (``smallish'') wave number. 
Within our model, we have now demonstrated 
that this result gives a valid long-time asymptotic for the 
kinematic dynamo problem. 
The key to understanding this coincidence lies in the fact that 
the magnetic-field correlation function decays exponentially
at large scales [asymptotic solution~\exref{sln_int_plus}].  
Due to the solenoidality of the magnetic field, 
such exponential tail implies 
that~$H(k)$ must vanish at~$k=0$ [see \apref{ap_sol_pot}, 
formula~\exref{spectrum_k4}]. Imposing this as a boundary 
condition on the small-scale spectrum~\exref{sln_K} 
necessarily implies~\exref{lambda_k0} or~\exref{deltalambda_n}.   

The solution we have thus obtained 
is relevant in three dimensions, where~$\lambda_0>0$. 
Note that formula~\exref{deltalambda_n} can be used to estimate 
the critical value of the Prandtl number that has to be 
exceeded in order for the dynamo effect to occur (i.e., for 
the growth rate to be positive, $\lambda=\lambda_0-\delta\lambda>0$):  
$\Pr_c \sim \exp\bigl(2\pi\sqrt{A/\lambda_0}\bigr)$. 
For the incompressible case, $\Pr_c\sim 26$; for the irrotational 
case, $\Pr_c\sim 17000$. It must be clear, of course, that 
such critical values 
are not to be taken as anything more than rough estimates, 
for they have been deduced from the formula~\exref{deltalambda_n}, 
which is strictly valid only when~$\delta\lambda\ll 1$. 
Numerical studies of the excitation threshold for the Kazantsev 
equation give (in the incompressible case) values of the 
critical Prandtl number (or critical magnetic Reynolds number, 
which, for a one-scale velocity field, is the same thing) 
about~$\Pr_c\sim53$ \citep{Novikov_Ruzmaikin_Sokoloff,Maslova_Ruzmaikin}. 

\paragraph{The case $\lambda_0<\lambda<0$ (damping).}
Here we must reconcile the relations \exref{match_gl} 
and \exref{match_c1c2} in the limit~$\yres/y_0\to0$. 
In view of the incompatibility of~\eqref{consist_gg},
this can only be done if  
\bea
\label{consist_gl}
\gamma_0\cot(\gamma_0 +\phi) - s_1 = 0.
\eea
This fixes the value of the phase~$\phi$ for any 
applicable~$\lambda$. No further constraint ensues 
for~$\lambda$ lying in the interval~$\lambda_0<\lambda<0$. 
Provided this interval is nonempty (as is the case in 
two dimensions, unless~$a=0$), a continuous spectrum 
of eigenvalues exists and fills the entire interval. 
Since the upper bound of the spectrum is zero, a power-like 
decay of the magnetic fluctuations should be expected. 

\paragraph{The case $\lambda<\lambda_0~\&~\lambda<0$ (damping).}
In this case, just like in the case of~$0\le\lambda<\lambda_0$, 
assuming that~$\delta\lambda$ remains finite (or becomes large) 
as~$y_0/\yres\to\infty$ 
leads to incompatibility between the right- and left-hand sides 
of~\eqref{match_ll}: on the right-hand side, there must be 
a finite constant, while the left-hand side has no definite 
limit. If~$\lambda_0>0$ (3D), this is the only possibility. 
Thus, {\em if~$\lambda_0$ is positive, negative values 
of~$\lambda$ are ruled out.} 

If~$\lambda_0\le0$ (2D), it is possible to have 
$\delta\lambda\to+0$ as~$y_0/\yres\to\infty$. Again, 
the argument that led to the solution~\exref{deltalambda_n} 
remains valid. However, this solution is not relevant 
because we have already discovered a continuous spectrum 
between~$\lambda_0$ and~$0$. It is the resulting power-like 
decay that will determine the long-time behavior 
of the magnetic fluctuations. 
The case of~$\lambda_0=0$, 
which occurs in~2D for~$a=0$ (the ``quasiequilibrium'' 
case we noted in~\ssecref{spreading}), 
appears to be exceptional 
in this context, for then the interval~$(\lambda_0,0)$ 
is empty and the only applicable solution is, 
in fact,~\exref{deltalambda_n} (where now~$\lambda<\lambda_0=0$). 
Namely, we have $\lambda \simeq -A\pi^2/\bigl[\ln(\Pr^{1/2})\bigr]^2\to -0$, 
which should also result in a power-like temporal decay law in the limit 
of vanishing resistivity ($\eta\to+0$).

\paragraph{Remark on the case~$\lambda=\lambda_0$.}
When~$\lambda=\lambda_0$, we have~$s_1=s_2=1/2$ and 
the solution~\exref{sln_inert} appears to be degenerate. 
This degeneracy is spurious. In fact, the power-like 
solutions~\exref{sln_inert} are asymptotics of the modified 
Bessel functions~$I_{\sqrt{(\lambda-\lambda_0)/A}}$
and~$K_{\sqrt{(\lambda-\lambda_0)/A}}$. This can be seen 
if the Kazantsev eigenvalue problem~\exref{Kaz_ev_prob} 
is solved in the inertial range with next-order correction 
terms in the Taylor expansions  of~$K(y)$ and~$Q(y)$ included 
[see~\eqref{sln_inert_corr} in~\apref{ap_next_order} below]. 
When~$\lambda=\lambda_0$, we should therefore use, instead of 
the asymptotic solution~\exref{sln_inert},
\bea
\label{sln_inert_zero}
\psi(y) = y^{1/2}\(c_1 + c_2\ln y\).
\eea 
The matching at~$\yres$ and~$y_0$ can be carried out in 
quite the same fashion as before. 
The consistency conditions as~$\ln(y_0/\yres)\to\infty$  
are given by~\eqref{consist_gg} if~$\lambda_0\ge0$ 
(still incompatible) and~\eqref{consist_gl} if~$\lambda_0<0$ 
(can always be satisfied).

\subsection{Discussion}
\label{discussion}

Thus, assuming that the particular form of the large-scale 
regularization of the Kazantsev quantum mechanics is unimportant, 
or, in other words, that the Kazantsev quantum-mechanical particle 
is ``localized'' at small (inertial) scales, 
we have been able to make progress 
in our investigation of the long-time asymptotic behavior 
of the passive magnetic fluctuations. In~2D, a continuous 
spectrum of negative~$\lambda$'s exists, which indicates 
a power-like temporal decay. Its exact nature 
can be discovered only by solving the time-dependent 
dynamo equation~\exref{eq_H_LL}. 
In~3D, a universal dynamo growth rate has been found, 
which asymptotically approaches~$\lambda_0\gamma$ from below, 
as~$\Pr\to\infty$ (the convergence is square-logarithmic).

The exact quantitative criterion that would clearly define the 
class of velocity correlators within which such universality 
with respect to the large-scale regularization 
exists, is unknown. Due to the complicated structure of the 
potential~\exref{potential}, it appears to be very hard to 
determine. On the other hand, no examples of 
realizable\footnote{I.e., those whose Fourier transforms are 
correlation functions (see~\apref{ap_iso_FT}).} 
velocity correlators 
for which the dynamo growth rate is essentially determined 
by the large-scale properties of the velocity field are available.
As we already mentioned in the Introduction,
numerical solution of the Kazantsev eigenvalue problem~\exref{Kaz_ev_prob} 
for various plausible forms of the velocity 
correlator \citep[see][and \apref{ap_next_order}]{Novikov_Ruzmaikin_Sokoloff,Artamonova_Sokoloff,Maslova_Ruzmaikin}, 
including the one corresponding to the Kolmogorov turbulent 
spectrum \citep{KA}, 
did not reveal any nonuniversal behavior of the second-order 
statistics of the magnetic field.  
Thus, the validity of the universal 
results presented in this section appears quite robust. 

A tentative explanation for such robustness could be given along 
the following lines. The mass~$m(y)$ of the Kazantsev particle 
at integral scales tends to the constant finite 
value~$m(\infty) \simeq 1/2\kappa_0$. 
At small (inertial) scales, however, 
it sharply increases and, at the resistive scales, reaches 
the asymptotically infinite value~$m(0)=1/4\eta$ (for $\eta\to+0$).
Thus, the particle is impelled to slide toward the small scales 
not just by the potential, but also by its own variable mass: 
it is ``heavier at small scales.'' 
This leads to 
the large-scale effects being marginalized
even if the value of the potential in the inertial range,~$V(y)\sim -V_0$, 
is not, in fact, its global (or even local) minimum. 
Note that, while the latter situation has not been encountered in 
the previous studies of the incompressible case, 
it turns out to be quite common when the advecting flow possesses 
some degree of compressibility (see \apref{ap_next_order}). 

It remains to be seen if examples of velocity correlators can be constructed 
for which the second-order statistics depart from 
the universal solutions obtained above. 
The issue of whether higher-order statistics may be 
more vulnerable in this respect is also open to investigation. 
At the current level of understanding of the two-point 
kinematic dynamo, we may conclude that, while the nonuniversality 
of the long-time solutions due to large-scale effects 
remains no less of a possibility than it was when first 
hinted at by \citet{Kazantsev}, the results obtained 
assuming small-scale universality appear quite reliable 
for practical purposes.


\section{Where Do We Go From Here?}
\label{sec_future}

The main physical conclusion of the kinematic theory developed above 
is that, at the moment when the nonlinear effects become important, 
the magnetic energy is concentrated at the smallest scales that 
it is able to reach at the exponential rate~$\sim\gamma$ during 
the kinematic stage of the dynamo. An ascending power 
spectrum~$\sim k^{3/2}$ extends through the magnetic inertial 
range up to the wavenumber where the peak of the spectrum is 
located at the time the nonlinearity sets in. 

It is important 
to understand that the second-order (spectral) theory developed 
here does not contain all of the statistical information 
necessary to determine the condition for the onset of the 
nonlinearity. Indeed, the nonlinear back-reaction effect is 
controlled by the Lorentz tension force~$\propto(\vB\cdot\nabla)\vB$ 
in the momentum equation of the fluid (in the incompressible case).
The second-order statistics of this quantity are fourth order 
in the magnetic field. Importantly, the statistics of~$(\vB\cdot\nabla)\vB$ 
are determined not just by the magnitude of the magnetic field 
and by its overall characteristic scale, but also by the {\em structure} 
of the field, namely, by the characteristic scale of its  
variation {\em along itself}. This issue was thoroughly investigated 
by Ott and coworkers \citep[see review by][]{Ott_review}, 
\citet{Cattaneo_review1,Cattaneo_review2}, and by \citet{SCMM_folding} 
(see also references therein). The result is that the parallel 
scale does not decay with the overall characteristic scale, 
but stays at values comparable to the velocity scale~$\kd^{-1}$. 
The small-scale field is arranged in a pattern of randomly oriented 
long (of characteristic size~$\sim\kd^{-1}$) folds, whithin which 
the field undergoes rapid transverse spatial oscillations of its 
direction. The implication is that the condition for the onset 
of the nonlinearity is the approximate equalization of the magnetic-field 
energy and the energy of the smallest turbulent eddies. 

Once this has occurred, a nonlinear theory is needed to describe futher 
evolution of the field. No such theory exists as yet. 
However, there does exist a fair amount of numerical evidence, 
which, along with the heuristic physical insight, can be used  
as guidance in mapping out the tentative strategies of picking 
up where the kinematic theory left off and building on its results. 
Resolving three main issues appears to be the first order of 
business at this point. 

\paragraph{Nonlinear saturation at subviscous scales.} 
The onset of the nonlinear regime can occur under two 
distinct sets of circumstances, depending on the magnitude 
of the initial seed field and on the magnetic Prandtl number. 
As we saw in the preceding sections, the growth of 
the field strength and the refinement of its scale proceed 
in parallel. With the seed field assumed to be concentrated 
at velocity scales~$\sim\kd^{-1}$, the time it takes 
for the dynamo to switch from the diffusion-free 
to the resistive regime is of order~$\sim\gamma^{-1}\log\Pr$. 
Given a very weak seed field, the dynamo will still be 
kinematic at the time of this transition. The input state 
for the nonlinear theory will then be that with the small-scale 
magnetic-energy spectrum extending through the entire 
subviscous scale range and peaked at the resistive scale~$\sim\kres^{-1}$ 
(see~\ssecref{spreading_diss}, \ssecref{quasist_k} and 
\secref{sec_corr_fns}). 
Alternatively, 
it is conceivable that the initial field should be sufficiently strong 
for the nonlinear regime to commence before the resistive scale 
is reached \citep{Kulsrud_lecture} (see~\ssecref{spreading}). 
One wonders then whether the 
nonlinear back reaction is capable of arresting all further 
refinement of the field scale. Although 
the answer to this question is in the negative, it appears 
that the spreading of the magnetic excitation into the 
remainder of the subviscous range available to it 
proceeds at a much slower pace than in the kinematic 
regime: most likely, at the resistive time scale 
\citep{Kinney_etal_2D,Kinney_etal_3D,SMCM_stokes}. 
It may be possible to study the evolution of the magnetic 
spectrum through this initial nonlinear stage in a 
theoretical framework 
that is substantially simpler than the full 
nonlinear~MHD. Namely, the action of the fluid motions 
on the small-scale magnetic fields can be modelled 
by balancing the viscous and magnetic stresses, rather than 
by solving the full Navier--Stokes equation with Lorentz 
back reaction (the so-called {\em Stokes model}). 
An ``intermediate'' small-scale nonlinear theory 
emerges from such an approach that is a direct extension 
of the kinematic theory elaborated above 
\citep{Kinney_etal_2D,SMCM_stokes}; an exactly solvable 
model of such a regime (without diffusion) 
was constructed by \citet{Boldyrev_alpha}. 

\paragraph{Plasma damping mechanisms.} 
A parallel question is how well the Spitzer resistive-diffusive 
small-scale regularization that is a part of the standard 
MHD paradigm describes what really happens at subviscous 
scales in the high-Prandtl-number astrophysical plasmas. 
It must be realized that, while the resistive 
scales~$\kres^{-1}$ that obtain by substituting the ISM or 
protogalactic-plasma parameters into the Spitzer formula 
for the magnetic diffusivity~$\eta$ may be 7~to~11~decades 
below the viscous scale~$\kd^{-1}$, the mean free path 
of the particles that make up these plasmas is substantially 
larger (the simplest estimate gives~$\lmfp\sim\Re^{-1/4}\kd^{-1}$, 
i.e., only about 10~times shorter than the viscous scale). 
The MHD approximation is not strictly valid at scales below the 
mean free path, and one must allow that a number of plasma damping 
mechanisms may supercede the Ohmic resistive 
dissipation \citep{Vainshtein_plasma,KA,Kulsrud_etal_report}. 
These include, most importantly, the ambipolar damping that 
is present due to the partial ionization of the ISM 
\citep{KA,Chandran_visc_rlx,Subramanian-a,Subramanian-b,Subramanian-c,Brandenburg_Subramanian} 
as well as the kinetic effects due to the \citet{Braginskii} tensor 
viscosity and magnetic diffusivity \citep{Malyshkin_phd} 
and to Landau damping of magnetic fluctuations \citep{Kulsrud_etal_report}. 
We note that, while 
all of these effects are nonlinear in the magnetic field, 
the kinetic effects depend solely on the direction, not the 
magnitude, of the field, which only needs to be sufficient to 
magnetize the ions --- a condition that is satisfied in the 
astrophysical plasmas in question already for very weak fields, 
which are otherwise passive. Treatment of these effects again 
appears possible by means of certain amendments to the kinematic 
theory developed here 
\citep[cf.][]{KA,Subramanian-a,Subramanian-b,Subramanian-c,SK_kinetic,Malyshkin_phd}. 
Note that, in view of the possibility of other-than-resistive  
small-scale regularization, the study of the onset of back reaction 
in the diffusion-free dynamo, as discussed in the previous 
paragraph, acquires additional importance.

\paragraph{Growth of large-scale magnetic fields.}
Finally, the most important --- and difficult to resolve --- issue 
is that of the feasibility of some form of nonlinear 
transfer of the magnetic energy accumulated at small scales 
toward large scales: {\em the inverse cascade}. 
Two distinct directions of inquiry must be clearly 
identified here. 

First, as we mentioned in~\ssecref{astro_motives}, it has long been 
known that the presence of a helical component in the fluid 
turbulence leads to exponential growth of the large-scale 
magnetic field in the kinematic regime ({\em the $\alpha$~effect}). 
In the high-Prandtl-number astrophysical plasmas, this growth 
is much slower than that of the small-scale fluctuations. 
Once the latter have entered the nonlinear regime, they are expected 
to substantially modify the $\alpha$~effect,  
almost certainly suppressing it to some extent. 
The magnitude of this suppression is a subject of 
an ongoing debate, various estimates ranging from extreme 
quenching with suppression factors as large as~$\Rm$ 
to moderate order-one reductions \citep[see, e.g, discussion in 
a recent series of papers by][and references therein]{Field_Blackman_Chou,Blackman_Field-a,Blackman_Field-b,Chou_alpha,Field_Blackman}. 
While none of the existing theories of the nonlinear 
$\alpha$~effect is universally agreed on, it appears safe 
to expect that 
(i) helicity does lead to some growth of the large-scale magnetic field;
(ii) this growth is slowed down in a nonlinear fashion 
by the accumulating small-scale magnetic fluctuations. 
At large Prandtl numbers, the suppression effect may be 
very large, making impossible any growth of the large-scale 
fields at time scales faster than the resistive. 
The kinematic $\alpha$-effect growth rate, which is already 
about 10000~times smaller than that of the small-scale dynamo, 
is thus further reduced by a factor possibly as large as~$\Rm$. 
In the context of the ISM's nearly perfect conductivity, such 
a dynamo would be quite prohibitively slow and could not explain the 
observed galactic magnetic field. 
[For some numerical evidence regarding these processes, 
we address the reader to papers by~\citet{Meneguzzi_Frisch_Pouquet,Cattaneo_Hughes_alpha,Brandenburg,Maron_Blackman}. 
However, none of these simulations has enough resolution 
to be conclusive with regard to the high-Prandtl-number plasmas.] 
  
The other avenue of investigation, which we announced  
in~\ssecref{astro_motives}, is to look for the possibility 
of nonhelical inverse cascade leading to eventual equipartition 
of the magnetic and kinetic energies at all scales including 
the large. Indeed, such an eventual equipartition 
appears eminently plausible at the first glance, 
since it corresponds to the steady-state spectra 
that follow from Kolmogorov-style theories based on the view 
of the MHD turbulence as resulting from nonlinear interactions 
of Alfv\'en waves \citep[][--- in this context, the outcome of the 
ongoing controversy over the exact form of the saturated 
equipartition spectrum is immaterial]{Iroshnikov,Kraichnan_MHD_spectrum,Goldreich_Sridhar-a,Goldreich_Sridhar-b}. 
However, several recent numerical simulations with~$\Pr>1$ 
and no externally imposed dc background field 
have cast serious doubt on the 
possibility of the saturated equipartition state 
\citep{Kinney_etal_2D,Kinney_etal_3D,Chou,Maron_Cowley,SMCM_stokes}. 
The questions of what the saturated magnetic spectrum is 
and whether the magnetic energy at the large scales can be 
shown to achieve values sufficient to give rise to an adequate 
theory of the observed galactic field, remain unresolved.  

\acknowledgments

We are grateful to J.~A.~Krommes who read an earlier manuscript 
of this work. Both the substance and the style of the presentation 
have benefited from his suggestions. 
We would also like to thank E.~Blackman, S.~Cowley, 
J.~Maron, V.~Pariev, V.~Rytchkov, D.~Uzdensky,
and the anonymous referee 
for useful comments. This work was supported in part by the 
U.~S.~Department of Energy under Contract~No.~DE-AC02-76-CHO-3073.

\appendix

\section{CORRELATION FUNCTIONS OF ISOTROPIC RANDOM FIELDS 
AND OTHER MATTERS}
\label{ap_review}

In this Appendix, we provide a set of useful facts and formulas 
that are frequently needed in turbulence calculations such as those 
of \Secref{sec_spectra} and \Secref{sec_corr_fns}. We will work with 
a random field $u^i$, which we will often refer to as the velocity 
field. However, all the results of this Appendix also hold for the 
magnetic field $B^i$ provided solenoidality constraint is properly 
implemented. We find it most convenient to work in $d$~dimensions. 
For~$d=3$, many of the results below can be found in the 
lucid and comprehensive chapter on the mathematical description of 
turbulence by \citet{Monin_Yaglom}.

This Appendix is organized as follows. In \ssecref{ap_iso_FT}, we provide 
the transformation formulas that relate the correlation functions 
of isotropic fields in configuration and wave-number (Fourier) spaces.
These impose certain constraints on the classes of functions in 
the $\vx$ space that are proper correlation functions.  
In \ssecref{ap_sol_pot}, the additional constraints arising from 
solenoidal and potential nature of the fields are explained. 
In \ssecref{ap_realizability}, the so-called realizability 
conditions are discussed that determine some essential features 
of the theories dealing with isotropic random fields. 
In \ssecref{ap_1to2pt}, we give some relations 
between the small-scale-expansion coefficients of 
the two-point correlation functions in the $\vx$~space 
and their counterparts in the $\vk$~space. 
These relations provide a bridge between the formalisms 
of \secref{sec_spectra} and \secref{sec_corr_fns} 
and are also needed to put the results of this paper and the 
one-point statistical results published earlier \citep{BS_metric,SK_tcorr,Boldyrev_tcorr,Boldyrev_alpha,SCMM_folding} 
in a unified theoretical framework. 
Finally, in~\ssecref{ap_angle_int}, we list the angle-integration 
identities that were used in the derivations of \secref{sec_spectra}.

\subsection{Isotropic Fourier Transforms}
\label{ap_iso_FT}

We will employ the usual Fourier-transform conventions:
\bea
\label{FT_direct}
u^i(\vk) &=& \intx\,e^{-i\vk\cdot\vx}u^i(\vx)
\qquad{\rm (direct),}\\
\label{FT_inverse}
u^i(\vx) &=& \intk e^{i\vk\cdot\vx}u^i(\vk)
\qquad{\rm (inverse).}
\eea
In this Appendix, we are not concerned with time dependence of our fields 
and concentrate purely on their properties arising from the assumptions 
of spatial homogeneity, isotropy, and mirror invariance.
Under this set of assumptions, we may define the correlation 
functions of the field~$u^i$ as follows: in the configuration space, 
\bea
\<u^i(\vx)u^j(\vx')\> &=& \kappa^{ij}(\vx-\vx'),\\
\label{kappa_x}
\kappa^{ij}(\vy) &=& \kappaLL(y)\delta^{ij} 
- \[\kappaLL(y)-{\kappaNN(y)\over d-1}\]\(\delta^{ij} - {y^iy^j\over y^2}\),
\eea
where $\kappaLL(y)$ and $\kappaNN(y)$ are called the {\em longitudinal} 
and the {\em transverse} correlation functions for obvious reasons.
In the Fourier space,
\bea
\label{homo_k}
\<u^i(\vk)u^j(\vk')\> &=& (2\pi)^d\delta(\vk+\vk')\kappa^{ij}(\vk),\\ 
\label{kappa_k}
\kappa^{ij}(\vk) &=& \kappa(k)\delta^{ij} + \tkappa(k)\,{k_i k_j\over k^2}.
\eea
The correlation tensors $\kappa^{ij}(\vy)$ and $\kappa^{ij}(\vk)$ 
defined above are Fourier transforms of one another:
\bea
\label{FT_kappa}
\kappa^{ij}(\vy) = \intk e^{i\vk\cdot\vy}\kappa^{ij}(\vk).
\eea
Let us substitute~\exref{kappa_k} into the above formula and find that
\bea
\label{kappa_vy}
\kappa^{ij}(\vy) = \delta^{ij}\intk e^{i\vk\cdot\vy}\kappa(k) 
- {\d^2\over\d y^i\d y^j}\intk e^{i\vk\cdot\vy}{\tkappa(k)\over k^2}. 
\eea 
The $d$-dimensional Fourier transforms of the isotropic (radial) functions 
can be found with the aid of the following handy formula 
\cite[see, e.g.,][]{Stein}:
\bea
\label{Phi_formula}
\int{\rm d}\Omega_d\,e^{i\vk\cdot\vy} = 
(2\pi)^{d/2}{J_{(d-2)/2}(ky)\over (ky)^{(d-2)/2}} 
= S_d\Phi_d(ky),
\eea
where ${\rm d}\Omega_d$ is the $d$-dimensional angular differential, 
and $S_d = \int{\rm d}\Omega_d = 2\pi^{d/2}/\Gamma(d/2)$ 
is the area of the unit sphere in $d$~dimensions.
We have defined the function 
\bea
\Phi_d(z) = \Gamma(d/2)\,{J_{(d-2)/2}(z)\over (z/2)^{(d-2)/2}},
\eea
which is the kernel of the ``isotropic Fourier transform''
and has many agreeable properties. Two of them will be particularly important
for us: 
$\Phi_d(z)$~is analytic, its Taylor expansion for small~$z$ being
\bea
\label{Phi_expand}
\Phi_d(z\to 0) = 1 - {z^2\over 2d} + {z^4\over8d(d+2)} + \cdots;
\eea
and $\Phi_d(z)$~satisfies the following differential recursion relation: 
\bea
\label{Phi_diff}
\({1\over z}\,{\d\over\d z}\)^n\Phi_d(z) = 
(-1)^n 2^{-n}\[\({d\over2}\)_n\]^{-1}\Phi_{d+2n}(z),
\eea
where $(d/2)_n = \Gamma(n+d/2)/\Gamma(d/2)$ is the Pochhammer symbol. 
Note also that
$\Phi_2(z) = J_0(z)$ and $\Phi_3(z) = (\sin z)/z$.

The isotropic Fourier transform is then 
\bea
\intk e^{i\vk\cdot\vy}\kappa(k) = \Idk\,k^{d-1}\Phi_d(ky)\kappa(k).
\eea
A similarly defined transform sometimes appears in the literature on special 
functions under the name of Bochner transform. 
After some straightforward manipulations, we obtain 
[from~\exref{kappa_x} and~\exref{kappa_vy} by way of \exref{Phi_diff}]
\bea
\label{FT_kappaLL} 
\kappaLL(y) = \Idk\,k^{d-1}\biggl[\Phi_d(ky)\bigl[\kappa(k) 
+ \tkappa(k)\bigr]
-{d-1\over d}\,\Phi_{d+2}(ky)\tkappa(k)\biggr],\\
\label{FT_kappaNN}
\kappaNN(y) = (d-1)\,\Idk\,k^{d-1}\[\Phi_d(ky)\kappa(k) 
+ {1\over d}\,\Phi_{d+2}(ky)\tkappa(k)\].\quad
\eea
We may also define the energy (spectrum) function 
\bea
I(k) &=& \kappa^{ii}(\vk) = d\kappa(k)+\tkappa(k),\\
\label{FT_E}
I(y) &=& \kappa^{ii}(\vy) = \kappaLL(y) + \kappaNN(y) 
= \Idk\,k^{d-1}\Phi_d(ky)I(k). 
\eea

The importance of such formulas as~\exref{FT_kappaLL}, \exref{FT_kappaNN}, 
and~\exref{FT_E}, 
lies in that they impose constraints on what functions in the 
configuration space may in fact be proper {\em correlation functions}. 
Thus, for example, 
since the $\vk$-space spectrum~$I(k)$ must clearly be positive 
for all~$k$ and be properly cut off at small scales, 
formula~\exref{FT_E} stipulates that only 
an isotropic Fourier transform of such a function can be an $\vx$-space 
correlation function of~$u^i$. 

Note also that the analyticity of the Fourier kernel~$\Phi_d(ky)$ 
[it possesses the Taylor expansion~\exref{Phi_expand} around the 
origin] is the reason for the analyticity of the velocity 
correlator that passively advected fields ``feel'' at small 
scales in the Batchelor regime.

\subsection{Correlation Functions of Solenoidal and Potential Fields}
\label{ap_sol_pot}

Two important special cases of the field $u^i$ are incompressible 
(solenoidal) and irrotational (potential) fields. 
In the Fourier space, we have
\bea 
\tkappa(k) &=& -\kappa(k)\qquad {\rm (solenoidal)},\\ 
\kappa(k) &=& 0\qquad {\rm (potential)}.
\eea 
In the $\vx$~space, solenoidality and potentiality impose differential 
relations between the longitudinal and the transverse correlation 
functions. Namely, 
\bea
\label{vonKarman_rln}
\kappaNN(y) &=& y\kappaLL'(y) + (d-1)\kappaLL(y)
\qquad {\rm (solenoidal)},\\
\label{Yaglom_rln}
\kappaLL(y) &=& {1\over d-1}\,\bigl[y\kappaNN'(y) + \kappaNN(y)\bigr]
\qquad {\rm (potential)}.
\eea
[The relation~\exref{vonKarman_rln} is known as the von~K\'arm\'an condition.]
The equivalence of the above relations in their $\vk$ and $\vx$~space forms 
can, of course, be easily demonstrated with the aid of 
the formulas~\exref{FT_kappaLL} and~\exref{FT_kappaNN}. 
 
We may summarize the properties of the solenoidal fields as follows
\bea
\kappa^{ij}(\vk) &=& {1\over d-1}\,\(\delta^{ij} - {k^ik^j\over k^2}\)I(k),\\
\kappa^{ij}(\vy) &=& \kappaLL(y)\delta^{ij} 
+{1\over d-1}\,y\kappaLL'(y)\(\delta^{ij} - {y^iy^j\over y^2}\),
\eea
where, from~\exref{FT_kappaLL} and~\exref{FT_E},
\bea
\label{FT_kappaLL_sol}
\kappaLL(y) &=& {1\over d}\,\Idk\,k^{d-1}\Phi_{d+2}(ky)I(k),\\
\label{FT_E_sol}
I(y) &=& {1\over y^{d-1}}\,{\d\over\d y}\,y^d\kappaLL(y) 
= \Idk\,k^{d-1}\Phi_d(ky)I(k).
\eea
These formulas are especially important because they must always hold 
for the correlation functions of the magnetic field, which {\em is} solenoidal.
For example, the formula~\exref{FT_kappaLL_sol} 
imposes a nontrivial (and somewhat different from that for the energy) 
constraint on the class of functions eligible to be 
longitudinal correlation functions of a solenoidal field~$u^i(\vx)$.

Note that, if $\kappaLL(y)$ decays at large scales ($y\to\infty$) 
faster than~$1/y^d$, then the formula~\exref{FT_E_sol} implies 
an often-used property of the energy function
\bea
\int_0^{\infty}{\rm d}y\,y^{d-1} I(y) = 0.
\eea
This means that $I(y)$~cannot remain positive for all~$y$, 
and that $I(k)$, which is the inverse Fourier transform of~$I(y)$, 
vanishes at~$k=0$. In view of the expansion properties 
of~$\Phi_d(ky)$ [see~\exref{Phi_expand}], the latter 
implies that the spectrum of a solenoidal field has the following 
scaling in the limit of small wave numbers:
\bea
\label{spectrum_k4}
k^{d-1}I(k) \sim k^{d+1}\quad{\rm as}\quad k\to 0.
\eea
In~3D, this implies the well-known $k^4$~infrared scaling 
of the spectra of solenoidal fields \citep{Monin_Yaglom}.

\subsection{The Realizability Inequalities}
\label{ap_realizability}

It is easy to see that the formula~\exref{homo_k}, which is a consequence 
of the homogeneity of space, can be recast in the following form:
\bea
\kappa^{ij}(\vk) = L^{-d}\<u^i(\vk)u^{j*}(\vk)\>, 
\eea
where $L^d$ is the volume of space (henceforth set to~1) 
and the star means complex conjugation.
Let us pick an arbitrary real unit vector $\vn$ and take a full 
convolution of $n_in_j$ with the above expression: 
\bea
n_in_j\kappa^{ij}(\vk) = \<|\vn\cdot\vu|^2\>,
\eea
whence, and by using~\exref{kappa_k}, we obtain, for all~$\vk$,
\bea
0 \le n_in_j\kappa^{ij}(\vk) \le \<|\vu|^2\> = \kappa^{ii}(\vk),
\quad\\
0 \le \kappa(k) + \tkappa(k)\,{(\vn\cdot\vk)^2\over k^2} \le 
d\kappa(k) + \tkappa(k).
\eea 
We may now write $(\vn\cdot\vk)^2 = k^2\cos^2\theta$, where 
$\theta$~is the (arbitrary) angle between~$\vk$ and~$\vn$.
Thus we arrive at the following {\em realizability theorem}:
for any value of~$\theta$ and any~$k$ the following two 
inequalities must hold: 
\bea
\kappa(k) + \tkappa(k)\cos^2\theta \ge 0,\quad\\
(d-1) \kappa(k) + \tkappa(k)\sin^2\theta \ge 0.
\eea
In practice, we will be mostly content to use just the particular 
cases of the above arising from setting~$\theta=0$ and~$\theta=\pi/2$:
\bea
\kappa(k) + \tkappa(k) \ge 0,\\
\kappa(k) \ge 0.\quad\quad
\eea 

Let us now derive the realizability constraints that operate in the 
configuration space. Consider the {\em two-point structure function} of 
the field~$u^i$: 
\bea
\bigl<\bigl[u^i(\vy)-u^i(0)\bigr]\bigl[u^j(\vy)-u^j(0)\bigr]\bigr> 
= 2\kappa^{ij}(0) - 2\kappa^{ij}(\vy),
\eea
take the full double dot product of the above expression 
with an arbitrary unit vector~$\vn$, and use the expression~\exref{kappa_x} 
for the correlation tensor~$\kappa^{ij}$. 
The resulting realizability theorem~is
\bea
\kappaLL(0) - \kappaLL(y)\cos^2\theta 
- {\kappaNN(y)\over d-1}\,\sin^2\theta 
= {1\over2}\bigl<\bigr|\vn\cdot\bigl[u^i(\vy)-u^i(0)\bigr]\bigl|^2\bigr>
\ge 0
\eea
for any~$y$ and any~$\theta$ (the angle between~$\vn$ and~$\vy$).
The inequalities of most practical value are again obtained 
for~$\theta=0$ and~$\theta=\pi/2$: 
\bea
\label{realizability_kappaLL}
\kappaLL(y) \le \kappaLL(0),\qquad\quad\quad\\
\kappaNN(y) \le (d-1)\kappaLL(0) = \kappaNN(0). 
\eea
Thus, the two-point correlation functions can never exceed 
the values they take when the two points coincide. 


\subsection{Small-Scale Expansion of Second-Order Correlation Functions}
\label{ap_1to2pt}

Since much of this work is concerned with one-point statistics, 
let us establish a correspondence between one- 
and two-point correlation properties of our field~$u^i$. 
For $y\to0$, we will routinely expand 
\bea
\label{corr_1pt}
\kappa^{ij}(\vy) = \kappa_0\delta^{ij} 
-{1\over 2}\,\kappa_2\(y^2\delta^{ij} + 2ay^iy^j\) 
+{1\over 4}\,\kappa_4y^2\(y^2\delta^{ij} + 2by^iy^j\) 
+\cdots.
\eea
Here $a$ and~$b$ are the so-called compressibility parameters 
which change within the intervals 
\bea
-{1\over d+1} \le a \le 1 \quad {\rm and} \quad
-{2\over d+3} \le b \le 2,
\eea
where the lower bounds correspond to the incompressible case, 
the upper bounds to the irrotational one.
Making use of the formula~\exref{Phi_expand}, we expand 
the expressions~\exref{FT_kappaLL} and~\exref{FT_kappaNN} for 
$\kappaLL(y)$ and $\kappaNN(y)$ at small~$y$, substitute them 
into~\exref{kappa_x}, and, by comparing with~\exref{corr_1pt}, 
establish the small-scale expansions of the velocity 
correlation functions in the $\vx$~space,
\bea
\kappaLL(y) &=& \kappa_0 - {1\over2}\,(1+2a)\kappa_2 y^2 
+ {1\over4}\,(1+2b)\kappa_4 y^4 + \cdots,\\
\kappaNN(y) &=& (d-1)\(\kappa_0 - {1\over2}\,\kappa_2 y^2 
+ {1\over4}\,\kappa_4 y^4 + \cdots\),
\eea
and the expressions for the parameters of the correlator~\exref{corr_1pt} 
in terms of the spectral characteristics of the velocity field,
\bea
\kappa_0 &=& {1\over d}\intk I(k),\\
\kappa_2 &=& {1\over d(d+2)}\intk k^{2}\bigl[(d+2)\kappa(k) + 
\tkappa(k)\bigr],\\
\kappa_4 &=& {1\over 2d(d+2)(d+4)}\intk k^4\bigl[(d+4)\kappa(k) + 
\tkappa(k)\bigr],\\
a &=& \kappa_2^{-1}{1\over d(d+2)}\intk k^{2}\tkappa(k),\\ 
b &=& \kappa_4^{-1}{1\over d(d+2)(d+4)}\intk k^{4}\tkappa(k),
\eea 
where the $\vk$-space integrations of radial functions could, of course, 
have been written more explicitly~as 
\bea
\intk = \Idk k^{d-1}.
\eea

Finally, let us remark that the expression~\exref{def_gamma} that was 
obtained in \secref{sec_spectra} for the initial growth rate of the total 
magnetic energy can be, in view of the above, written as follows:
\bea
\label{gamma_1to2pt}
\gamma = {1\over2}\,(d-1)(da+2)\kappa_2 = 
{1\over2}\,{d-1\over d+1}\(d + 2 + \igamma\)\kappa_2,
\eea 
where~$\igamma = d[1+a(d+1)]$. The above expression 
is naturally in agreement with the results known 
in the one-point setting \citep[see, e.g.,][]{BS_metric,SK_tcorr}.
Another convenient expression (or definition) for~$\gamma$, 
that we use in \secref{sec_corr_fns}, is 
\bea
\label{gamma_kA}
\gamma = {1\over2A}\,(1+2a)\kappa_2 = {|\kappaLL''(0)|\over2A},
\eea
where $A$~is defined by the formula~\exref{def_A}.

\subsection{Angle-Integration Identities}
\label{ap_angle_int}

Dealing with $d$-dimensional wave-number integrations 
while deriving 
the mode-coupling equation~\exref{MC_eq}, 
the evolution laws~\exref{eq_W} and~\exref{eq_msk}, 
and the SSF~equation~\exref{SSF_eq}, 
required repeated use of some simple angle-integration identities 
which we list here for completeness:
\bea
\int{\rm d}\Omega_d &=& S_d = {2\pi^{d/2}\over\Gamma(d/2)},\\
\int{\rm d}\Omega_d\,n_i n_j &=& {S_d\over d}\,\delta_{ij},\\
\int{\rm d}\Omega_d\,n_i n_j n_k n_l &=& 
{S_d\over d(d+2)}\(\delta_{ij}\delta_{kl} 
+ \delta_{ik}\delta_{jl} + \delta_{il}\delta_{kj}\),\\
\int\int{\rm d}\Omega_d{\rm d}\Omega'_d (\vn\cdot\vn')^2 &=& 
{S_d^2\over d},\\
\int\int{\rm d}\Omega_d{\rm d}\Omega'_d (\vn\cdot\vn')^4 &=& 
{3 S_d^2\over d(d+2)}, 
\eea
where ${\rm d}\Omega_d = {\rm d}\phi\sin\theta_1{\rm d}\theta_1\dots
\sin^{d-2}\theta_{d-2}{\rm d}\theta_{d-2}$ is the $d$-dimensional 
angular differential, 
and $\vn=\vk/k$ is a unit vector in the direction of~$\vk$. 
The above identities can be derived in various ways, one of them 
being via formula~\exref{Phi_formula}.

\section{THE FURUTSU--NOVIKOV FORMULA}
\label{ap_Novikov}

Here we state an extremely useful Gaussian-averaging theorem 
that is widely used throughout this work. This result is due 
to \citet{Furutsu} and \citet{Novikov} and is 
generally referred to as the Furutsu--Novikov, 
or ``Gaussian-integration,'' formula. 

Consider a random Gaussian vector field~$u^i(q)$, 
where~$q$ is the aggregate of all variables it depends  
on (such as time, spatial position, Fourier variable, or vector index). 
Let~$R[\vu]$ be some functional of the field~$u^i$. 
Then the following holds:
\bea
\<u^i(q) R[\vu]\> = \int\diff q'\<u^i(q)u^j(q')\>
\<{\delta R[\vu]\over\delta u^j(q')}\>,
\eea
where the integration is over all possible values of~$q'$.

This property of Gaussian fields is simply a generalization 
of the well-known ``splitting'' property of the Gaussian 
averages.

\section{LARGE-SCALE CORRECTIONS TO THE INERTIAL-RANGE ASYMPTOTICS}
\label{ap_next_order}

In this Appendix, we work out the first ``large-scale'' 
corrections to the inertial-range asymptotics~\exref{m_inert} 
and~\exref{V_inert} of the Kazantsev particle's mass 
and the potential it lives in, as well as the appropriate 
corrected asymptotic forms of the solutions. 
These corrections are 
needed in order for us to get a better idea of 
the general form of the Kazantsev quantum mechanics. 
They were also appealed to when the asymptotic matching 
was performed in~\ssecref{matching}.

In order to obtain the desired corrections, 
we must use the expansion~\exref{corr_1pt} of the velocity 
correlator~\exref{xi_correlator_x} up to the fourth order in~$y$.  
The fourth-order Taylor expansions of the renormalized 
diffusivities are 
\bea
K(y) &\simeq& 2\eta + {1\over2}\,(1+2a)\kappa_2 y^2 
- {1\over4}\,(1+2b)\kappa_4 y^4,\\
Q(y) &\simeq& 2\eta + {1\over2}\,\kappa_2 y^2 
- {1\over4}\,\kappa_4 y^4,
\eea
and we find the following corrected asymptotics for 
mass and potential in the inertial~range:
\bea
\label{mass_4}
m(y) &\simeq& {1\over(1+2a)\kappa_2 y^2}
\[1 + {1\over2}\,{1+2b\over1+2a}\,{\kappa_4\over\kappa_2}\,y^2\],\\
\label{potential_4}
V(y) &\simeq& -V_0 + V_2 y^2,\\ 
&&\where
V_2={3\over8}\[3d - 7 - 4b - (d-3)^2{1+4a-2b\over6(1+2a)^2}\]\kappa_4.
\nonumber
\eea

\begin{figure}
\plotone{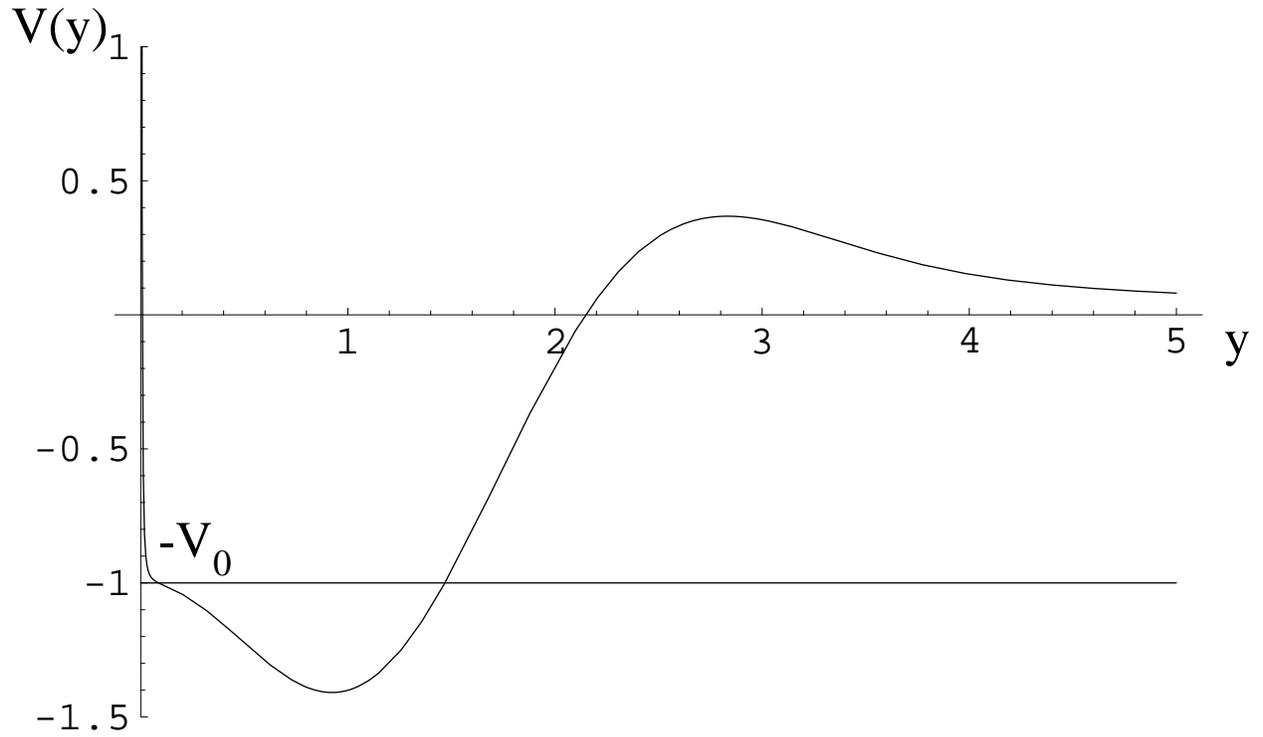}
\caption{The Kazantsev potential in the case of 3D~irrotational 
flow. For reference, the line corresponding to~$V=-V_0$ is also shown.
$V(y)$~is measured in units of~$V_0$, $y$~in units 
of~$(\kappa_0/\kappa_2)^{1/2}$. The transverse velocity correlation 
function was chosen in the 
form~$\kappaNN(y)=2\kappa_0\exp\(-\kappa_2 y^2/2\)$. 
Due to the irrotational property of the flow, 
the longitudinal correlation function had to be calculated 
according to the formula~\exref{Yaglom_rln} of~\apref{ap_sol_pot}.
\label{fig_V_irr}}
\end{figure}

The correction~$V_2 y^2$ in the formula~\exref{potential_4} 
is positive in the~3D~incompressible case, 
changes sign at~$b=1/2$, and remains negative 
for all~$b>1/2$, including in the irrotational case. 
This means that the value~$V(y)\simeq-V_0$ represents 
an intermediate ``plateau,'' and not the bottom of the potential well, 
which extends further down [how far down is, of course, 
impossible to determine on the basis of the Taylor 
expansion~\exref{potential_4}; a representative example 
of such a potential is given in~\figref{fig_V_irr}]. 
The danger thus arises that the Kazantsev particle might 
slide into this deeper recess and thereby destroy 
the small-scale universality, upon which hinged 
the validity of the results obtained in~\ssecref{matching}. 
Whether or not this happens 
in reality may depend on the large-scale structure of 
the potential and on the interplay between it 
and the structure of the variable mass~$m(y)$. The importance 
of the latter effect is clear from the form of the eigenvalue 
problem~\exref{Kaz_ev_prob} itself, which can be reinterpreted 
as a Schr\"odinger equation for a particle with {\em constant mass} 
and {\em zero energy} in a $\lambda$-dependent 
effective potential~$U_\lambda(y)$:
\bea
\label{eq_with_U}
-\psi'' + U_\lambda(y)\psi = 0,\qquad\\
U_\lambda(y) = 2m(y)\bigl[V(y) + \lambda\gamma\bigr].
\eea
We must now treat~$\lambda$ as a parameter 
to be chosen in such way that the zero-energy eigenstate exist 
and be a ground state. The effective potential~$U_\lambda(y)$ explicitly 
incorporates both influences [variable mass and potential~$V(y)$] that 
act on the Kazantsev particle. In the inertial range, 
it is, of course, {\em an inverse-square potential.} 
Due to the inverse-square decay of~$m(y)$, the fourth-order effects 
lead only to a constant correction, namely,
\bea
\label{U_lambda}
U_\lambda(y) &\simeq& -{\lambdamax-\lambda\over Ay^2} + U_2,\\
&&\where
U_2 = {1\over\gamma A}\[V_2 - 
{1\over4}\,(1+2b)\kappa_4\,{\lambdamax-\lambda\over A}\].
\nonumber
\eea
The solutions of~\eqref{eq_with_U} with the effective 
potential~\exref{U_lambda} are the modified Bessel functions: 
\bea
\label{sln_inert_corr}
\psi(y) = y^{1/2}\[
c_1 I_{\sqrt{(\lambda-\lambda_0)/A}}\bigl(U_2^{1/2}y\bigr) 
+ c_2 K_{\sqrt{(\lambda-\lambda_0)/A}}\bigl(U_2^{1/2}y\bigr)\].
\eea
These are the solutions we referred to in justifying 
the asymptotic~\exref{sln_inert_zero} of~\ssecref{matching}.  

\begin{figure}
\plotone{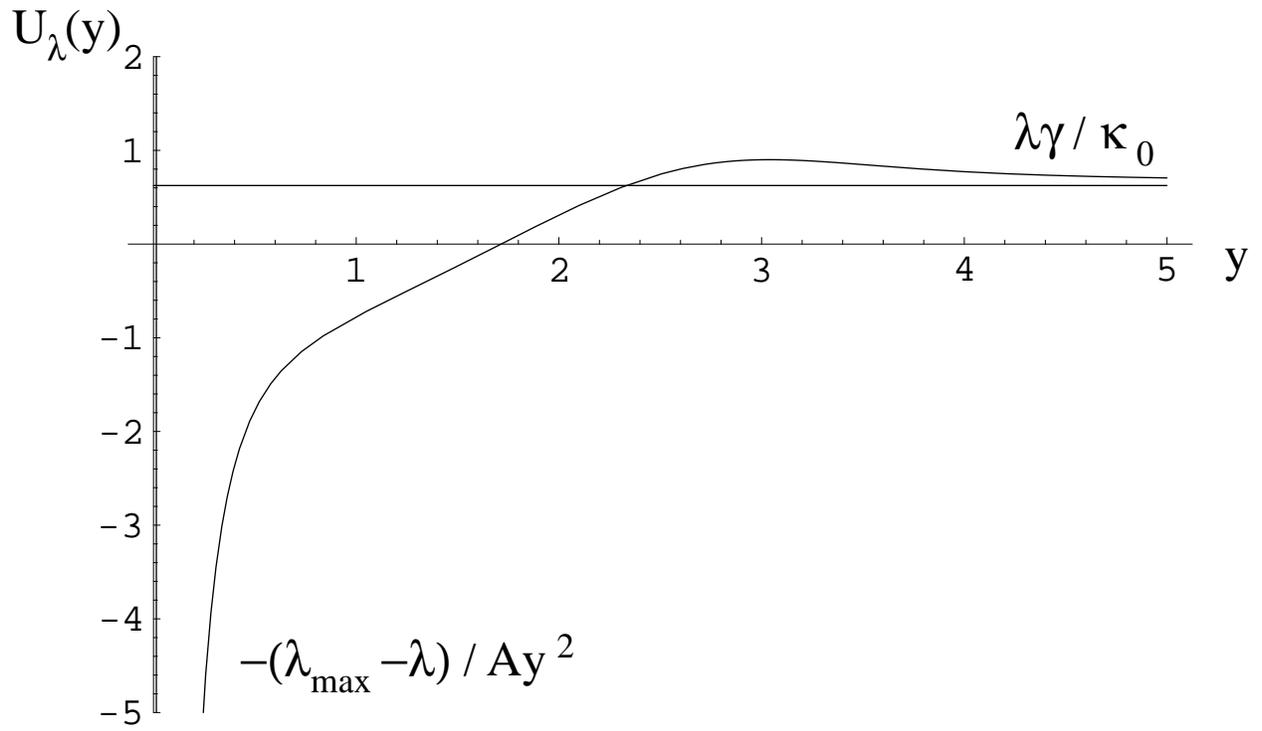}
\caption{The form of the 
function~$U_\lambda(y)=2m(y)\bigl[V(y)+\lambda\gamma\bigr]$ which  
corresponds to the potential~$V(y)$ plotted in~\figref{fig_V_irr} 
and~$\lambda=\lambda_0=1/8$. 
The bottom of the potential well extends to~$U_\lambda\sim-\kappa_2/\eta$ 
and is not shown. 
\label{fig_U_irr}}
\end{figure}

The potentials~$V(y)$ and~$U_\lambda(y)$ for a particular 
3D~irrotational velocity field 
with the transverse correlation function chosen to be 
$\kappaNN(y)=2\kappa_0\exp\(-\kappa_2 y^2/2\)$, 
are shown in~\figref{fig_V_irr} 
and~\figref{fig_U_irr}. In this specific case, the straightforward 
numerical solution of the eigenvalue problem~\exref{Kaz_ev_prob} 
proved that the Kazantsev particle was ``heavy enough'' to settle 
on the flat ``porch'' corresponding to the inertial-range 
asymptotic~$V(y)\simeq-V_0$, rather than slide off towards larger 
scales and deeper into the potential well. 
The small-scale universality therefore persisted. 

\placefigure{fig_V_irr} 
\placefigure{fig_U_irr}
\notetoeditor{These two figures should appear side-by-side in print}

\clearpage

\clearpage

\clearpage

\clearpage

\clearpage

\clearpage

\clearpage

\clearpage

\clearpage

\clearpage

\clearpage

\begin{deluxetable}{llll}
\tabletypesize{\scriptsize}
\tablewidth{0pt}
\tablecaption{Physical parameters in the Galaxy and 
Protogalaxy.\label{tab_params}}
\tablehead{
\colhead{Parameter} & 
\colhead{Notation} &
\colhead{Galaxy (warm ISM)} & 
\colhead{Protogalaxy}}
\startdata
\cutinhead{Densities and Temperatures}
neutral (hydrogen) density, cm$^{-3}$ & $n_n$ &
$1$ & $0$ \\ 
ion (proton) density, cm$^{-3}$ & $n_i$ &
$1$ & $10^{-3}$ \\ 
temperature, K & $T\sim T_i \sim T_e \sim T_n$ &
$10^4$ & $10^6$\\
\cutinhead{Length Scales, cm 
($3\cdot10^{18}$ cm $\approx$ 1 parsec)} 
system size & --- &
$10^{22}$ & $10^{24}$ \\
magnetic field coherence & --- &
$10^{21}$ & {\rm unknown} \\ 
energy-containing & $L$ &
$10^{20}$ & $10^{24}$ \\
viscous cutoff & $\kd^{-1}\sim\Re^{-3/4}L$ &
$10^{16}$ & $10^{20}$\\
neutral mean free path & $\ell_n\sim\vth\min\{\tau_{nn},\tau_{ni}\}$ & 
$10^{14}$ & ---\\
ion mean free path & $\ell_i\sim\vth\min\{\tau_{ii},\tau_{in}\}$ &
$10^{12}$ & $10^{19}$\\
resistive cutoff & $\kres^{-1}\sim\Pr^{-1/2}\kd^{-1}$ &
$10^{9}$ & $10^{9}$\\
ion skin depth & $d_i\sim c/\omega_{pi}$ & 
$10^{7}$ & $10^{9}$\\ 
\cutinhead{Time Scales, sec 
($\pi\cdot10^{7}$ sec $\approx$ 1 year)}
system lifetime & --- &
$10^{17}$ & $10^{17}$\\
system rotation & --- &
$10^{15}$ & ---\\
largest-eddy turnover & $L\vth^{-1}$ &
$10^{14}$ & $10^{17}$\\ 
viscous/smallest-eddy turnover & $\teddy\sim(\kd\ueddy)^{-1}$ &
$10^{12}$ & $10^{14}$\\ 
neutral-neutral collision & 
$\tau_{nn} \sim4\pi r_{\rm Bohr}^2\vth n_n$ &
$10^{9}$ & ---\\ 
ion-neutral/neutral-ion collision & 
$\tau_{in}\sim(m_in_i/m_nn_n)\tau_{ni}$ &
$10^{8}$ & ---\\ 
ion-ion collision & 
$\tau_{ii} \sim(k_B T)^{3/2}m_i^{1/2}(e^4n_i\ln\Lambda)^{-1}$ &
$10^{6}$ & $10^{12}$\\ 
\cutinhead{Velocities, cm/sec}
thermal/shock/largest-eddy & 
$\vth \sim(k_B T/m_i)^{1/2}$ &
$10^6$ & $10^7$ \\ 
smallest-eddy & $\ueddy\sim\Re^{-1/4}\vth$ & 
$10^5$ & $10^6$\\ 
\cutinhead{Viscosities, cm$^2$/sec} 
neutral viscosity & $\nu_n\sim\vth^2\tau_{nn}$ &
$10^{21}$ & ---\\
ion viscosity& $\nu_i\sim\vth^2\tau_{ii}$ &
$10^{18}$ & $10^{26}$ \\
magnetic (Spitzer) diffusivity & 
$\eta \sim(k_B T)^{-3/2}m_e^{1/2}e^2c^2\ln\Lambda/4\pi$ &
$10^{7}$ & $10^{4}$\\
\cutinhead{Dimensionless Numbers}
hydrodynamic Reynolds & $\Re\sim\vth L/\nu$ &
$10^5$ & $10^4$ \\
magnetic Reynolds & $\Rm\sim\vth L/\eta$ & 
$10^{19}$ & $10^{27}$\\
magnetic Prandtl & $\Pr\sim\nu/\eta\sim(\kres/\kd)^2$ &
$10^{14}$ & $10^{22}$ \\
\enddata
\end{deluxetable}

\clearpage

\begin{deluxetable}{cccccccccc}
\tablewidth{0pt}
\tablecaption{Parameters of the SSF equation.\label{SSF_params}}
\tablehead{
\colhead{Dimension} & 
\colhead{Velocity Field} & 
\colhead{$a$} & 
\colhead{$A$} & 
\colhead{$B$} & 
\colhead{$C$} &
\colhead{$s_0$} & 
\colhead{$\xi_0$} & 
\colhead{$\lambda_0$} & 
\colhead{$\lambdamax$}}
\startdata
$d=3$     & Incompressible & -1/4 &
1/5  & 2/5  & 4/5 & -1/2  & 3/2     & 3/4         & 4/5\\
          & Irrotational   &  1   &
3/10 & 3/5  & 1/5 & -1/2  & 3/2     & 1/8         & 1/5\\
$d=2$     & Incompressible & -1/3 &
1/4  & -1/4 & 0   &  1    & 2       & -1/4        & -3/16\\
          & Irrotational   &  1   &
3/4  &  5/4 & 0   & -1/3  & 2/3     & -1/12       & 5/48\\  
\enddata
\end{deluxetable}

\end{document}